\documentclass{aa}  
\usepackage{graphicx}
\usepackage{multirow}
\usepackage{txfonts}
\usepackage{hyperref}

\usepackage[dvipsnames]{xcolor}
\usepackage{caption,subcaption} 
\usepackage{float}

\begin{document}

   \title{GASTLI}

   \subtitle{An open-source coupled interior-atmosphere model to unveil gas giant composition}

   \author{L. Acu\~{n}a\inst{1}
          \and
          L. Kreidberg\inst{1}
          \and
          M. Zhai\inst{1,2}
          \and
          P. Mollière\inst{1}
          }

   \institute{Max-Planck Institut für Astronomie, Königstuhl 17, 69117 Heidelberg, Germany\\
              \email{acuna@mpia.de}
         \and
        Chinese Academy of Sciences South America Center for Astronomy (CASSACA), National Astronomical Observatories, Chinese Academy of Sciences, Beijing 100101, China}
        
   \date{Received 30 April 2024; accepted --}

 
  \abstract
   {The metal mass fractions of gas giants are a powerful tool to constrain their formation mechanisms and evolution. The metal content is inferred by comparing mass and radius measurements with interior structure and evolution models. In the midst of the JWST, CHEOPS, TESS, and the forthcoming PLATO era, we are at the brink of obtaining unprecedented precision in radius, age and atmospheric metallicity measurements.
   To prepare for this wealth of data, we present the GAS gianT modeL for Interiors (GASTLI), an easy-to-use, publicly available Python package. The code is optimized to rapidly calculate mass-radius relations, and radius and luminosity thermal evolution curves for a variety of envelope compositions and core mass fractions. Its applicability spans planets with masses $17 \ M_{\oplus} < M < 6 \ M_{Jup}$, and equilibrium temperatures $T_{eq} < 1000$ K.
   The interior model is stratified in a core composed of water and rock, and an envelope constituted by H/He and metals (water). The interior is coupled to a grid of self-consistent, cloud-free atmospheric models to determine the atmospheric and boundary interior temperature, as well as the contribution of the atmosphere to the total radius.
   We successfully validate GASTLI by comparing it to previous work and data of the Solar System’s gas giants and Neptune. We also test GASTLI on the Neptune-mass exoplanet HAT-P-26 b, finding a bulk metal mass fraction between 0.60-0.78 and a core mass of 8.5-14.4 $M_{\oplus}$. Finally, we explore the impact of different equations of state and assumptions, such as C/O ratio and transit pressure, in the estimation of bulk metal mass fraction. These differences between interior models entail a change in radius of up to 2.5\% for Jupiter-mass planets, but more than 10\% for Neptune-mass. These are equivalent to variations in core mass fraction of 0.07, or 0.10 in envelope metal mass fraction.}
   


   \maketitle
%

\section{Introduction}

Gas giants ($M > 20 \ M_{\oplus}$) consist mainly of H / He, with a small percentage of metals (water, rock). The total amount of metal is sensitive to the formation process, likely either core accretion or gravitational instability. In the former, pebbles and planetesimals cluster together to constitute a compact core of a few Earth masses (up to 10 $M_{\oplus}$). Then, the gravitational pull of this core is enough to cause the runaway accretion of metal-poor H/He gas, until the protoplanetary disk dissipates \citep{HM21}. Core accretion is the most widely accepted process to explain the trend observed in bulk metal content with planet mass, known as the mass-metallicity relation, observed for both Solar System gas giants and exoplanets \citep[][and references therein]{HM21,Helled23, Thorngren16,Teske19}. On the other hand, gravitational instability is a better explanation for the existence of very massive gas giants ($M > 1 \ M_{Jup}$), especially at large orbital distances, and around M-dwarfs \citep{Mercer20}. Additionally, an enrichment in metals in gas giants is difficult to explain in-situ formation, which suggests that planetesimal accretion impacts the bulk composition during inward migration \citep{HM21}.

The bulk metal content of exoplanets cannot be inferred from density directly, as this is not the result of interior composition alone, but also of mass, irradiation (or equilibrium temperature), and age. Therefore, interior structure models are required to disentangle the effects of metal content from the other planetary parameters on density. These models provide mass-radius relations as well as the thermal evolution of luminosity and radius with planetary age \citep{Burrows93,Burrows95,Burrows97,Burrows2001,Marley12,Marley21}, which can be used as the forward models for retrievals on mass, radius, age and atmospheric metallicity, if available \citep{Dorn15,Thorngren16,Otegi20,Muller21,Bloot23}. 
Estimates of bulk metal content are limited to warm and cold gas giants ($T_{eq}$ < 1000 K). The radii of hotter planets are inflated, possible due to tidal heating \citep{Bodenheimer01,Leconte10,Ibgui10,Ibgui11}, or Ohmic dissipation \citep{Batygin10,Perna10}. The advection of potential temperature is a third mechanism that can contribute to the inflation of hot Jupiters, which consists on a source of heat due to high-entropy fluid parcels in the deep atmosphere, caused by a strong irradiation and the atmosphere's longitudinal momentum conservation \citep{Tremblin17,Sarkis21,Schneider22}. However, the exact inflation mechanism is not well constrained, and it is difficult to fine-tune its effect in interior models, which introduces a degeneracy between bulk composition and inflation extent. This is normally parameterized with an extra heat free variable \citep{Thorngren18}.

In the following, we summarize codes and tools that are publicly available to carry out interior modelling analysis and retrievals. For gas giants, \cite{Fortney07,Thorngren16} (F07) and \cite{Muller21} (MH21) provide mass-radius-age tables for interpolation that span a wide range of masses and irradiations, although their envelope metallicities are fixed to solar for the former, while the latter cannot model planets whose envelope metallicity is equal or greater than the planet's core mass fraction. This limits the parameter space of compositional parameters, especially for planets whose envelope metallicity has been probed via atmospheric characterization, and it is different from solar. For Neptune-mass planets, the mass-radius relations by \cite{Zeng19} are widely used. However, these models assume that the envelope is a  pure H/He layer, and simplify the calculation of the entropy in the interior-atmosphere boundary, yielding very different surface temperatures from self-consistent models in radiative-convective equilibrium \citep{Rogers23}. Instead, \cite{LF14} (LF14) are more appropriate to infer the metal content of Neptune-mass planets. Similarly to the F07 models, LF14 fix their atmospheric composition to solar, although an enhancement in metallicity for more metal-rich envelopes is discussed. Finally, there are currently two interior structure models that are open source: MESA \citep{Paxton11,Paxton13,Paxton15,Paxton18,Paxton19} and MAGRATHEA \citep{magrathea}. The former is a widely used stellar interior structure model, that has been adapted to be applied to gas giants \cite{Muller21} and sub-Neptunes \cite{chen_rogers16}. Nonetheless, its implementation in Fortran and computational times per model ($\sim$ 1-3 minutes) make it difficult to couple with detailed non-gray atmospheric models. In contrast, MAGRATHEA is a computationally fast interior model for low-mass planets, although its layer structure (ideal gas EOS on top of liquid/ice water layer) poses a challenge to its adaptation for more massive planets with non-ideal, high pressure H/He envelopes for new users.

Given the limitations of the models described above, there is a need for an easy-to-use, flexible interior model applicable to a wide range of exoplanet masses and compositions. Such a tool would not only enable the interior structure and composition of newly discovered planets by observers, but would also allow theorists to test the effect of different assumptions on mass-radius and thermal evolution calculations. For example, differences in equations of state (EOS) and thermodynamical data can impact significantly the inferred bulk metal mass \citep{Chabrier19,HG23}. Other assumptions such as the determination of the temperature in the outer interior boundary or radiative-convective boundary with different atmospheric models \citep{Burrows97,Baraffe2003,Chen23} and clear/cloudy atmospheres \citep{Poser19,Poser24} can also impact the radius and evolution obtained by interior models. Furthermore, it would also enable homogeneous analyses of large samples in the context of exoplanet populations \citep{Thorngren16,Acuna22,Baumeister23}.

In this work, we present the GAS gianT modeL for Interiors (GASTLI), a coupled interior and atmosphere model applicable to warm ($T_{eq} < 1000$ K), Neptune-mass and gas giant planets ($17 \ M_{\oplus} < M < 6 \ M_{Jup}$) with a wide variety of envelope compositions (0.01 $\times$ solar < [Fe/H] < 250 $\times$ solar, and 0.1 < C/O < 0.55) and core mass fractions (0 to $\sim$0.99). GASTLI is a user-friendly, open-source Python package\footnote{\url{https://github.com/lorenaacuna/GASTLI}}\footnote{\url{https://gastli.readthedocs.io/en/latest/}}. In Sect. \ref{sec:interior}, \ref{sec:atm_models} and \ref{sec:coupling}, we describe the interior structure and atmosphere models, as well as their self-consistent coupling. In Sect. \ref{sec:thermal_evol}, we indicate how a sequence of interior models at different internal temperatures are used to determine the radius and luminosity evolution at a constant composition, mass and irradiation. GASTLI is validated by comparing it to previous work and data for the Solar System gas giants and Neptune (see Sect. \ref{sec:jupiter} and \ref{sec:Neptune}). We also showcase the applicability of GASTLI to exoplanets in Sect. \ref{sec:exoplanets}, where we analyse the bulk metal content of the well-characterised Neptune-mass HAT-P-26 b, and display the evolution of the mass-radius relations with age. Finally, we discuss the impact of different input EOS data and assumptions in the calculations of mass-radius relations and thermal evolution and summarize our conclusions in Sect. \ref{sec:discussion} and \ref{sec:conclusion}, respectively.


\section{Interior structure model} \label{sec:interior}

\begin{figure*}[h]
   \centering
   \includegraphics[width=0.7\hsize]{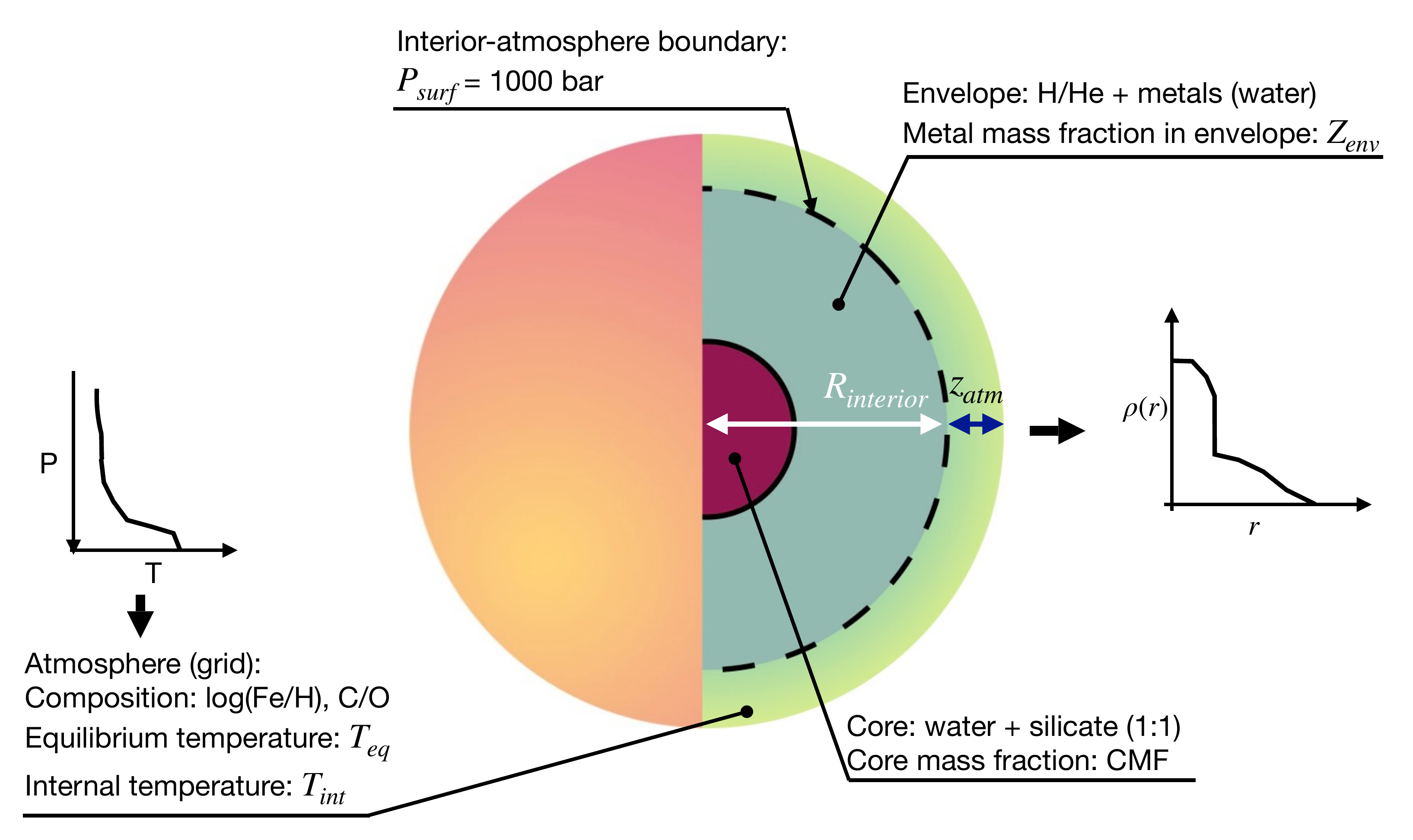}
      \caption{Schematic figure showing the layers, composition and parameters of our interior structure model, GASTLI. The parameters set by the user are the planet mass, core mass fraction (CMF), envelope composition, equilibrium temperature (at null albedo) and internal temperature. The envelope composition is either determined by log(Fe/H) in solar units or as an envelope metal mass fraction, $Z_{env}$. The outputs computed by the interior structure model include the total planet radius, $R = R_{interior} + z_{atm}$, and the interior profiles (see text).}
         \label{fig:interior_diagram}
\end{figure*}

In this section, we present our interior structure model, shown schematically in Fig. \ref{fig:interior_diagram}. To self-consistently calculate the boundary conditions and the contribution of the atmosphere to the total radius and planetary profiles, we coupled the interior model to a grid of atmospheric models. We describe in detail our one-dimensional (1D) radiative-convective atmospheric models in Sect. \ref{sec:atm_models}.

\subsection{Physical model} \label{sec:basics_interior}
In the 1D interior structure model, the one-dimensional grid represents the radius of the planet, from its center up to the outer interface of the outermost layer. For simplicity, we will refer to this boundary as the surface, even though gas giants do not present a physical surface like terrestrial planets do. The planetary interior is stratified in two layers, which are the core and the envelope. In gas giants, the pressure is high enough for water and rock to be mixed homogeneously \citep{Vazan22}. We therefore assume a 1:1 mixture of rock and water for the core. Furthermore, the envelope is a mixture of H/He and water, where the mass fraction of water in the envelope corresponds to the input parameter $Z_{env}$. Hence, the composition of the interior model is controlled by two parameters: the core mass fraction (CMF), and the mass fraction of water (representative of metals) in the envelope. The total envelope mass fraction (EMF) can be calculated as EMF = 1 - CMF, while the mass fraction of H/He in the envelope corresponds to $x_{H-He} = X_{env}+Y_{env}$ = $1-Z_{env}$. $X_{env}$ and $Y_{env}$ are the mass fractions of hydrogen and helium in the envelope, respectively.

The interior structure profiles, which are the pressure, temperature, gravity acceleration, density and entropy are calculated at each point of the radial grid. The pressure, $P(r)$, is computed according to hydrostatic equilibrium (Eq. \ref{eq:hydro_eq}); the gravitational acceleration, $g(r)$, is governed by Gauss's theorem (Eq. \ref{eq:gauss_eq}). For the temperature, we assume that the main heat transport mechanism is convection, and therefore the temperature profile is adiabatic (Eq. \ref{eq:adiabat_eq}). In these differential equations (Eq. \ref{eq:hydro_eq} - \ref{eqn:gruneisen}), the parameter 
$m$ corresponds to the mass at radius $r$, and $G$ is the gravitational constant. The variable $\gamma$ is the Gr\"uneisen parameter, which determines the vibrational properties given then size of the lattice of the crystalline structure of a material. The Gr\"uneisen parameter relates the temperature of the crystal and the density, $\rho(r)$, obtained with the Equation of State (EOS). In addition, the internal energy $E$ and volume $V$ are also required to calculate the Gr\"uneisen parameter. Finally, $\phi$ is the seismic parameter, which is computed as the derivative of the pressure with respect to the density.

In sum, the interior model solves the following set of equations:
\begin{equation} \label{eq:hydro_eq}
    \frac{dP}{dr} = - \rho g
\end{equation}

\begin{equation} \label{eq:gauss_eq}
    \frac{dg}{dr} = - 4 \pi G \rho - \frac{2 G m}{r^{3}}
\end{equation}

\begin{equation} \label{eq:adiabat_eq}
    \frac{dT}{dr} = - g \frac{\gamma T}{\phi}
\end{equation}

\begin{equation}
\label{eqn:gruneisen}
\begin{cases}
\phi = \dfrac{dP}{d \rho}  \\
\gamma = V \  \left(  \dfrac{dP}{dE} \right)_{V} 
\end{cases}
\end{equation}

\begin{equation}
\label{eqn:mass_conserv}
\dfrac{dm}{dr} = 4 \pi r^{2} \rho
\end{equation}

The boundary condition for Eq. \ref{eq:gauss_eq} is that the gravity at the center of the planet is zero, $g(r=0)=0$. On the other hand, the boundary conditions for the pressure and the temperature are their respective values at the surface interface, $P(r=R_{interior})=P_{surf}$ and 
$T(r=R_{interior})=T_{surf}$. $P_{surf}$ has a fixed value of 1000 bars, while $T_{surf}$ is determined by the atmospheric models and coupling algorithm (see Sect. \ref{sec:atm_models} and \ref{sec:coupling}). The mass of the planet and its two layers are computed assuming conservation of mass (Eq. \ref{eqn:mass_conserv}). 

We solve these differential equations by integrating them individually from the center of the planet outwards in each point of the spatial grid that represents the radius. We carry out the integration by the trapezoidal rule. Then we iterate the four profiles (pressure, gravity, temperature and density) sequentially until their differences from iteration to iteration are less than a precision of $10^{-5}$, or a maximum number of iterations of 30. In addition to these requirements, convergence is reached once the surface boundary conditions are satisfied, and the input mass of the planet is fulfilled. For more details, see \cite{Brugger16,Brugger17,Brugger_phd_thesis}.

\subsection{Mixtures} \label{sec:mixtures}
The composition of the core is 50\% water and 50\% silicates in mass. The envelope is constituted by H/He and metals. Since the exact composition of the metals in the deep envelope is unknown, we assume they are pure water in the interior model. The mass fraction of metals in the envelope is $Z_{env}$. This is defined by the user as an input parameter to the interior model. Thus, the EOS and thermodynamical properties of H/He, water and silicates are required to calculate the density and Gr\"uneisen parameter of the core and envelope mixtures.

To calculate the density of a mixture, we use the additive law of density (Eq. \ref{eqn:density_add}), where $\rho_{i}$ are the densities of the mixture end-members, and $x_{i}$ their respective mass fractions \citep{Pebbles64,Chabrier92,Nettelmann08}. In the case of the core, the two components are water and silicate, both with mass fractions of 0.50; whereas for the envelope, the end-members are H/He and water, with mass fractions of ($1-Z_{env}$) and $Z_{env}$, respectively. 

\begin{equation}
\label{eqn:density_add}
\dfrac{1}{\rho_{mix}} = \sum_{i=1}^{N} \frac{x_{i}}{\rho_{i}}
\end{equation}

Similar to the density, the Gr\"uneisen parameter of the individual components are employed to calculate the Gr\"uneisen parameter of the mixture. \cite{Duval71} find that the mass weighted Gr\"uneisen parameter differs appreciably from experimental values. Instead, it is recommended to assume the additive volume rule for the internal energy (Eq. \ref{eqn:energy_add}) or the Gibbs free energy. The additive volume rule can be substituted in the formal definition of the Gr\"uneisen parameter \citep[Eq. \ref{eqn:grun_formal}]{Arp84,Aguichine21} to show that the Gr\"uneisen parameter of a mixture is the weighted inverse mean of the end-members (Eq. \ref{eqn:grun_add}).

\begin{equation}
\label{eqn:energy_add}
u_{mix} = \sum_{i=1}^{N} x_{i} \ u_{i}
\end{equation}

\begin{equation}
\label{eqn:grun_formal}
\gamma = V \ \left( \frac{\partial P}{\partial U} \right)_{V} = \left(\frac{\partial P}{\partial u} \right)_{V}
\end{equation}

\begin{equation}
\label{eqn:grun_add}
\dfrac{1}{\gamma_{mix}} = \sum_{i=1}^{N} \frac{x_{i}}{\gamma_{i}}
\end{equation}

The entropy at the surface boundary, $S(r=R_{interior})=S_{top}$, is required to compute the thermal evolution of the planet (see Sect. \ref{sec:thermal_evol}). The entropy of a mixture is defined by the following law \citep{Baraffe08}: 

\begin{equation}
\label{eqn:entropy_add}
S_{mix} = \sum_{i=1}^{N} x_{i} \ S_{i} + S_{ideal}
\end{equation}

\noindent where $S_{ideal}$ is the contribution due to the ideal entropy of mixing. This extra term is taken into account for the H-He interactions (see Sect. \ref{sec:eos}). The contribution due to interactions between metals and H/He is dependent on the mean molecular weight and mean charge of the metal component \citep{Baraffe08}. In this work, we assume pure water to represent metals. However, the exact metals that are present in the envelope are largely unconstrained. Thus, we do not include the H/He-metal contribution due to the degeneracy in the different possible compositions of the ices. 


\subsection{Equations of state (EOS) and thermodynamical data} \label{sec:eos}

\subsubsection{H/He} \label{sec:hhe_eos}
We make use of the EOS tables provided by \cite{Chabrier21} (CD21) to compute the density and thermodynamical properties of H/He. CD21 take into account an ideal mixture correction for the density and the entropy at a He mass fraction $\tilde{Y}$ = 0.245 for the H-He mixture. However, the cosmogonic helium mass fraction we consider is $Y^{*}$ = 0.275. We use the correction terms tabulated by \cite{HG23} to subtract the correction at $\tilde{Y}$ = 0.245 and add the term at $Y^{*}$ = 0.275. The correction term, $\Delta V$, depends on the mass fraction of H, $X = 1-Y$, and He, $Y$; and the tabulated value, $V_{mix}$ \citep[Eq. \ref{eqn:tab_corr}]{HG23}. Thus, the corrected density can be expressed as a sum of the ideal density and the correction $\Delta V$ (Eq. \ref{eqn:rho_corr}). Hence, the density tabulated by CD21 can be expressed as the a mixture at $Y^{*}$ = 0.275 corrected with the term $Y^{*}$ = 0.275 (Eq. \ref{eqn:rho_CD21}).

\begin{equation}
\label{eqn:tab_corr}
\Delta V = X Y \ V_{mix}
\end{equation}

\begin{equation}
\label{eqn:rho_corr}
\frac{1}{\rho_{corr}} = \frac{1}{\rho_{ideal}} + \Delta V
\end{equation}

\begin{equation}
\label{eqn:rho_CD21}
\frac{1}{\rho_{CD21}} = \frac{1}{\rho_{ideal}(X^{*},Y^{*})} + \Delta V(\tilde{X},\tilde{Y})
\end{equation}

If we substitute Eqs. \ref{eqn:rho_CD21} and \ref{eqn:tab_corr} in Eq. \ref{eqn:rho_corr}, we obtain the corrected density at $Y^{*}$ = 0.275 as:

\begin{equation}
\label{eqn:rho_final}
\frac{1}{\rho_{corr}} = \frac{1}{\rho_{CD21}} + V_{mix} \ (X^{*} Y^{*} - \tilde{X} \tilde{Y})
\end{equation}

We follow a similar approach to correct the entropy consistently with the cosmogonic helium mass fraction, which yields:

\begin{equation}
\label{eqn:rho_final}
S_{corr} = S_{CD21} + S_{mix} \ (X^{*} Y^{*} - \tilde{X} \tilde{Y})
\end{equation}

\noindent where $S_{mix}$ is the correction tabulated by \cite{HG23}.

We compute the density and entropy profiles of a Jupiter analog for three cases: the original CD21 EOS, the EOS corrected at $Y^{*}$ = 0.275, and the EOS with no corrections. The densities corrected for non-linear mixing effects (the original CD21 and the $Y^{*}$ = 0.275 correction) agree with each other within less than 1\%. However, not including the non-ideal mixing effects can produce differences of up to 8\% in the density and entropy profiles. This agrees with \cite{HG23}, who report differences of up to 15\% in their models for Jupiter. This entails a difference of $\sim$ 0.20 $R_{\oplus}$, which is approximately 2\% of Jupiter's radius. Hence, we adopt the correction for $Y^{*}$ = 0.275 from \cite{HG23}. 

The Gr\"uneisen parameter for H/He is calculated using the thermodynamic quantities listed in the CD21 tables. \cite{Chabrier19} define this variable as:

\begin{equation}
\label{eqn:chabrier_grun}
\gamma = \frac{P}{\rho \ T \ C_{v}} \chi_{T}
\end{equation}

\noindent $C_{v}$ is the specific heat at constant volume, and $\chi_{T}$ is the isothermal compressibility. The specific heat capacities at constant volume and pressure are defined in Eqs. \ref{eqn:c_v} and \ref{eqn:c_p}, respectively. 

\begin{equation}
\label{eqn:c_v}
C_{v} = C_{p} - \frac{P}{\rho \ T} \frac{\chi_{T}^{2}}{\chi_{\rho}}
\end{equation}

\begin{equation}
\label{eqn:c_p}
C_{p} = S \ \left( \frac{\partial log S}{\partial log T} \right)_{P}
\end{equation}

The entropy, $S$, and the partial derivative in Eq. \ref{eqn:c_p} are tabulated by CD21. The compressibility at constant temperature (isothermal) and at constant density are shown in Eqs. \ref{eqn:chi_t} and \ref{eqn:chi_p}, respectively. The two partial derivatives necessary to compute the compressibilities are tabulated in the CD21 tables, completing the ensemble of thermodynamical variables required to evaluate Eqs. \ref{eqn:chabrier_grun} and \ref{eqn:c_v}.

\begin{equation}
\label{eqn:chi_t}
\chi_{T} = \left. \left( \frac{\partial log \rho}{\partial log T} \right)_{P} \ \middle/ \ \left( \frac{\partial log \rho}{\partial log P} \right)_{T} \right.
\end{equation}

\begin{equation}
\label{eqn:chi_p}
\chi_{P} = \left. 1 \ \middle/ \ \left( \frac{\partial log \rho}{\partial log P} \right)_{T} \right.
\end{equation}

\subsubsection{Water} 
To calculate the density and entropy of water, we use the subroutine provided by \cite{Mazevet19}. This function also estimates the internal energy, which together with the pressure, enables us to calculate the discrete derivative in Eq. \ref{eqn:grun_formal}. This EOS consists of a fit to experimental data obtained by the International Association for the Properties of Water and Steam \citep{Wagner02} for the supercritical phase of water, and quantum molecular dynamics simulations for plasma and superionic phases \citep{French09}. The validity of this water EOS spans to a maximum temperature of $10^{5}$ K, and a maximum density of 100 $g \ cm^{-3}$ \citep{Mazevet19}. Hence, its validity range makes it an adequate EOS for modelling the interior of warm gas giants, where temperatures of 10000 K, and densities of 25-30 $g \ cm^{-3}$ are expected \citep{Mankovich20,Neuenschwander21}.

\subsubsection{Silicates} 

We adopt the SESAME EOS for dry sand for the density and the thermodynamic properties of silicate rock \citep{sesame, Miguel22}. The Gr\"uneisen parameter is computed using Eqs. \ref{eqn:chabrier_grun} to \ref{eqn:chi_p} together with the entropy, density and partial derivatives of the quantities tabulated by the SESAME database. These tables span from 500 K to 20000 K in temperature, and reach a maximum pressure of 20 $g \ cm^{-3}$, having a validity range that covers the conditions in the deep interior of warm gas giants with masses lower than 1 $M_{Jup}$ \citep{Miguel22}. For exoplanets with $M > 1 \ M_{Jup}$, the thermodynamic quantities are evaluated at T = 20000 K when the temperature in the deep interior exceeds the maximum limit at which data are available.

\section{Atmospheric model} \label{sec:atm_models}

\subsection{Grid of self-consistent 1D models} 

To calculate the surface boundary temperature required for the interior model, we create a grid of self-consistent, 1D atmospheric models. The grid covers a range of surface gravities, equilibrium temperatures, internal temperatures, metallicities, C/O ratios (summarized in Table~\ref{tab:atm_grid}). The atmospheric models were generated with the Pressure–Temperature Iterator and Spectral Emission and Transmission Calculator for Planetary Atmospheres (\textit{petitCODE}) code \citep{Molliere15,Molliere17}. \textit{petitCODE} assumes a 1D plane-parallel approximation, and uses the correlated-k method to perform the radiative transfer calculations. We assume that the emission flux is globally averaged. To obtain the pressure-temperature profile, \textit{petitCODE} solves for the atmospheric thermal structure iteratively, enforcing radiative-convective and chemical equilibrium. In each iteration, the Schwarzschild criterion is evaluated to determine whether an atmospheric layer is convective or radiative. The model is considered to have converged once the maximum absolute difference (evaluated layer-by-layer) between the current iteration's temperature and the temperature 60 iterations ago is less than 0.01 K. In addition, the emerging flux at the top of the atmosphere must have a relative difference with the imposed flux of less than 0.001. The imposed flux is defined by the planetary effective temperature, $T_{\rm eff}$:

\begin{equation} \label{eqn:F_Teff}
    F_{\rm imposed} = \sigma T_{\rm eff}^{4}
\end{equation}

\noindent where $\sigma$ is the Stefan-Boltzmann constant. The effective temperature is computed from the internal temperature, $T_{\rm int}$, the equilibrium temperature, $T_{\rm eq}$, and the Bond albedo $A$ of the current atmospheric iteration:

\begin{equation} \label{eqn:Teff_sum}
    T_{\rm eff}^{4} = T_{\rm int}^{4} + (1-A)T_{\rm eq}^{4}
\end{equation}

In each iteration, the bond albedo $A$ is determined by comparing the scattered stellar flux at the top of the atmosphere to the incoming stellar flux. The atmospheric Bond albedo cannot be given as an input parameter, since it must be computed with the spectral energy distribution (SED) of the host star and the planetary atmospheric structure obtained by \textit{petitCODE}. We assume a Sun-like star for the stellar SED. The internal (or intrinsic) temperature is defined such that the outgoing flux at the top of the atmosphere is $F_{int} = \sigma T_{int}^4$ in the absence of external (stellar) irradiation. In the presence of external irradiation our definition of the effective temperature in Eq. \ref{eqn:Teff_sum} allows us to calculate the imposed outgoing flux at the top of the atmosphere as Eq. \ref{eqn:F_Teff}. 

The internal temperature parameterizes the net heat flux coming through the interior-atmosphere upwards from the deep interior. This heat flux $F_{int} = \sigma T_{int}^4$ encloses the secular heat flux resulting from the thermal cooling and contraction of the planet, plus other extra heat sources, such as tidal heating \citep{Leconte10,Agundez14}. The luminosity associated to the net heat flux is $L = L_{sec} + L_{extra}$ \citep{Poser19}. The secular luminosity is calculated as $L_{sec} = 4 \pi \sigma R_{pl}^{2} T_{thermal}^{4}$, while the luminosity caused by an extra heat source such as tidal heating can be expressed as $L_{extra} = 4 \pi \sigma R_{pl}^{2} T_{tidal}^{4}$. Since both luminosities have a similar factor of $4 \pi \sigma R_{pl}^{2}$, the user can simply provide the net internal temperature as $T^{4}_{int} = T^{4}_{thermal} + T^{4}_{tidal}$ to GASTLI if tidal heating is to be taken into account.


As the equilibrium temperature provides the incident stellar flux at the top of the atmosphere ($\sigma T_{eq}^{4}$), it depends on the star-planet parameters:

\begin{equation} 
    T_{\rm eq}^{4} = \frac{T_{\star}^{4}}{f_{\rm av}} \  \left( \frac{R_{\star}}{a_{d}} \right)^{2}
\end{equation}

\noindent where $a_{d}$ is the orbital semi-major axis; and $T_{\star}$ and $R_{\star}$ are the stellar effective temperature and radius, respectively. The factor $f_{\rm av}$ corresponds to $f_{\rm av}$ = 4 if the equilibrium temperature is assumed to be a global average, while $f_{\rm av}$ = 2 for the day-side average. In both of these cases \textit{petitCODE} assumes an isotropic incidence of the stellar light at the top of the atmosphere.

For the radiative transfer calculations, we take into account collision-induced absorption (CIA) and line absorption species. The continuum opacity sources are CIA due H$_{2}$-H$_{2}$ and H$_{2}$-He collisions. The line opacity absorbers include H$_{2}$O, CH$_{4}$, CO$_{2}$, HCN, CO, H$_{2}$, H$_{2}$S, NH$_{3}$, OH, C$_{2}$H$_{2}$, PH$_{3}$, Na, K. The references we use for the CIA, and atomic and molecular line opacities can be found in \cite{Molliere15,Molliere17}. We also include Fe, Fe+, Mg and Mg+ \citep{Kurucz93}, and TiO and VO (B. Plez, priv. comm.) as line absorbers.

 
The wavelength resolution we use to carry out the radiative transfer calculations is $R = \lambda / \Delta \lambda$ = 10. This resolution is high enough to determine the bolometric fluxes accurately in radiative-convective equilibrium calculations in the k-correlated method. \textit{petitCODE} separates the wavelength range into 80 bins from 0.11 to 250 µm, and resolves the cumulative opacity distribution function on 36 points in each bin, ensuring a proper resolution of the opacity's line cores. This approach was found to accurately reproduce the moment-averaged opacities and the Rosseland-mean and Planck-mean opacities, even at low pressures, all of which are needed when solving for the temperature using the Variable Eddington Factor method \citep{Molliere15,Baudino17}. To perform fast calculations of the self-consistent PT profiles, we pre-compute the mixed opacity tables for different sets of atmospheric abundances before the iterations start. These tables are sampled in a grid of 40 x 40 pressure and temperature points. Thus, the opacities are interpolated in each iteration from the pre-computed tables. The resolution in pressure-temperature space we chose is high enough to yield profiles that are consistent with those obtained by calculating the opacity tables in each iteration \citep{Molliere15}.


Given a list of reactant species, and the atomic number abundances, we assume equilibrium chemistry to obtain the molecular abundances and mass fractions. We employ easyCHEM\footnote{\url{https://easychem.readthedocs.io/en/latest/}} \citep{Molliere17}, which implements a Gibbs minimization scheme following the method presented for the Chemical Equilibrium with Application (CEA) code \citep{Gordon_cea,Mcbride_cea}. Among the reactant species, there are several species for which equilibrium condensation is taken into account. These include H$_{2}$O, VO, MgSiO$_{3}$, Mg$_{2}$SiO$_{4}$, Fe, Al$_{2}$O$_{3}$, Na$_{2}$S, KCl, and TiO$_{2}$.


Our atmospheric grid has two tables: the first one is for the temperature profile as a function of pressure (P-T profile), and the second one for the metal mass fraction as a function of pressure (P-$Z_{\rm atm}$ profile). The grid presents 6 dimensions, whose corresponding physical parameter, array size, scale and values are shown in Table \ref{tab:atm_grid}. The grid contains 5 $\times$ 10 $\times$ 10 $\times$ 5 $\times$ 2 = 5000 atmospheric models. Given its range of surface gravities at a pressure level of $P_{\rm surf}$ = 1000 bar, the grid of atmospheric models enables the coupled interior-atmosphere model to be applicable to planets with masses between $\sim$15 $M_{\oplus}$ and 7 $M_{\rm Jup}$.
In terms of envelope composition, the models span from sub-solar metallicity (Fe/H = 0.01 $\times$ solar), which is an almost pure H/He atmosphere, to Fe/H = 250 $\times$ solar, which is equivalent to $\sim$80\% metals in mass fraction. The C-to-O ratios extend from a low value, C/O = 0.10, to the solar ratio, C/O = 0.55. We assumed a Sun-like host star, with effective stellar temperature $T_{\star}$ = 5777 K, across the complete grid.

The P-T and P-$Z_{\rm atm}$ tables are interpolated by using Python's \verb|scipy.interpolate.RegularGridInterpolator|. This function requires regular axes to constitute a regular grid in hyperparameter space \citep{Virtanen20}. Thus, if a model is not available for a set of parameters in the grid, it is filled with  \verb|numpy.nan| values. This approach is computationally faster than having an irregular grid whose axes sizes are not constant.

\begin{table*}[h]
\centering
\caption{Physical parameters and their array values that constitute the axes of the grid of atmospheric models.}
\label{tab:atm_grid}
\begin{tabular}{lc}
\hline \hline
Parameter                       & Array values \\ \hline
log-Surface gravity, $log(g_{surf})$ [cm/s$^{2}$] & 2.60, 3.00, 3.39, 3.78, 4.18            \\
Equilibrium temperature, $T_{eq}$ [K] & from 100 to 1000 K, with $\Delta T_{eq}$ = 100 K          \\
Internal temperature, $T_{int}$ [K] & from 50 to 950 K, with $\Delta T_{int}$ = 100 K       \\
log-Metallicity, $log$(Fe/H) [$\times$ solar] & -2.0, 0.0, 1.0, 2.0, 2.4  \\
C/O ratio & 0.10, 0.55 \\
Pressure, $P$ [bar] & 130 points in log-scale from 1000 to 10$^{-6}$ bar      \\ \hline
\end{tabular}
\end{table*}

\subsection{Atmospheric profiles and thickness} \label{sec:atm_thick}

We calculate the radius profile by solving Eq. \ref{eqn:dpdm}, where the gravity $g(r)$ is computed with Eq. \ref{eqn:g_atm}. In these, $r$ is the planet radius and $m(r)$ is the enclosed mass.

\begin{equation}
\label{eqn:dpdm}
\frac{dP}{dr} = - g(r) \rho
\end{equation}

\begin{equation}
\label{eqn:g_atm}
g(r) = G \frac{m(r)}{r^{2}}
\end{equation}

We define the boundary condition for the atmospheric pressure and radius at $m = M_{interior}$, which is the mass enclosed between the planet center and the interior's surface interface. $R(m=M_{interior}) = R_{interior}$, and $P(m=M_{interior}) = $ 1000 bar. We assume that the observable transiting radius is located at 20 mbar \citep{Grimm18,Mousis20}. Thus, the total radius is obtained by evaluating the atmospheric radius profile, $r$ at a pressure $P$ = 20 mbar.

The atmospheric density profile is computed by evaluating the EOS for H/He and water along the pressure-temperature profile interpolated in the atmospheric grid's P-T table. We use the same EOS for H/He as the interior model (Sect. \ref{sec:hhe_eos}). In contrast, we adopt the AQUA EOS \citep{Haldemann20} for the density of metal species in the atmosphere, which are represented by water. The AQUA EOS is a collection of different EOS, in which each EOS is applied in its validity range of pressures and temperatures. The EOS used for water in the interior model, \cite{Mazevet19}, is applicable for low pressures, but only at temperatures below 1273 K. The two EOS contained in AQUA at low pressures and at temperatures relevant for the atmospheres of warm gas giants are \cite{Wagner02} and CEA \citep{Gordon_cea,Mcbride_cea}. Since \cite{Haldemann20} establish the transition from \cite{Mazevet19} EOS to these two other EOS at $P \sim$1000 bar, the coupling between the density in the interior model and the atmospheric model is smooth. We take into account the condensation of water in the density by setting it equal to the density at the saturation pressure, $\rho_{sat} = \rho(P = P_{sat})$, when $P \geq P_{sat}(T)$. The saturation curve function, $P_{sat}(T)$, is calculated by using the fit to laboratory data provided by \cite{Wagner02}.

Similar to the deep interior density, the atmospheric density is calculated by using the additive law (Eq. \ref{eqn:density_add}), with a metal mass fraction $Z_{atm}$ and (1-$Z_{atm}$) for water and H/He, respectively. If the user provides the atmospheric metallicity in $\times$ solar units, GASTLI interpolates the P-$Z_{atm}$ table in the grid of atmospheric models to obtain the metal mass fraction as a function of pressure. Then the input envelope metal mass fraction for the interior model, $Z_{env}$, is equal to this $Z_{atm}$ profile evaluated at $P$ = 1000 bar to keep the interior and atmosphere metal mass fraction consistent with each other. If the user provides as input $Z_{atm}$ directly, the atmospheric and deep envelope metal mass fraction are kept equal to this value. To interpolate the temperature in the P-T table, we convert from the input $Z_{atm}$ to $\rm log$(Fe/H) (see Table \ref{tab:atm_grid}) with Eqs. \ref{eqn:fortney_ratio} and \ref{eqn:fortney13} \citep{Fortney13}:

\begin{equation}
\label{eqn:fortney_ratio}
{\rm Fe/H} = \frac{(O:H)_{atm}}{(O:H)_{\odot}} 
\end{equation}

\noindent We adopt the solar value for the O-to-H ratio, $(O:H)_{\odot}$ = 4.89 $\times 10^{-4}$, from \cite{Lodders03}. The atmospheric O-to-H ratio is defined as: 

\begin{equation}
\label{eqn:fortney13}
(O:H)_{atm} = \frac{\frac{Z_{atm}}{X^{'}}(\mu_{X} / \mu_{Z})}{s + 2 \left[\frac{Z_{atm}}{X^{'}} (\mu_{X} / \mu_{Z}) \right]}
\end{equation}

\noindent where $X^{'}$ is the overall hydrogen mass fraction, and $\mu_{X}$ and $\mu_{Z}$ are the mean molecular weights for hydrogen and metals, respectively. The hydrogen mass fraction, $X^{'}$, does not include He. Thus, knowing that $Y^{*}$ = 0.275, we obtain that $X^{'} = (1-Z_{atm})/1.379$. We adopted water as the species representative of metals, so their molecular weight is $\mu_{Z} = \mu_{H2O} = 18$ g/mol. If hydrogen is in molecular form, $H_{2}$, the constant $s$ = 2, and $\mu_{X} = 2 \ \mu_{H} = 2$ g/mol, whereas for atomic hydrogen, $s$ = 1, and $\mu_{X} = 1 \ \mu_{H} = 1$ g/mol. For simplicity, we assume that most of the hydrogen is in molecular form in the upper atmosphere of warm gas giants in Eq. \ref{eqn:fortney13}. The difference in radius produced between the P-$Z_{atm}$ table and Eqs. \ref{eqn:fortney_ratio} and \ref{eqn:fortney13} is 0.005 $R_{Jup}$ for a Jupiter analog. This is due to a difference in temperature of 10-20 K at 1000 bar, having a negligible effect in the coupled interior-atmosphere model.


\section{Interior-atmosphere coupling} \label{sec:coupling}

\subsection{Algorithm} 

The temperature at the outermost interface (named surface in Sect. \ref{sec:basics_interior}), is an input parameter to the interior model, $T_{surf}$. In addition, the interior structure model also requires the planetary mass and composition (core mass fraction and envelope metal mass fraction) to compute the planet's interior radius, $R_{interior}$. This radius corresponds to the distance between the center of the planet and the interior's surface, located at a pressure $P$ = 1000 bar. On the other hand, the grid of atmospheric models take as input the planet's surface gravity, $g_{surf} = M/R_{interior}^{2}$, to interpolate the temperature at $P$ = 1000 bar. Consequently, the inputs of the interior model are not the direct outputs of the atmospheric models, and vice versa. This means their coupling is not straightforward. 


We use an iterative algorithm, previously presented in \cite{Acuna21}, to couple the interior model and the grid of atmospheric models self-consistently. In the following, we summarize the main structure of the algorithm for consistency. The algorithm starts by assuming an initial guess for the interior radius of the planet, $R_{guess}$, which can be adapted by the user. This is then used to obtain the surface gravity, and inputted to the grid of atmospheric models to have a first estimate of the surface temperature, $T_{surf}$. Subsequently, this value is used as the boundary condition of the interior model, which computes the interior radius, $R_{interior, \ i}$, for iteration $i$ = 0. Then if the condition $\vert R_{interior, \ i} - R_{guess} \vert$ is less than a given tolerance, the algorithm has converged to a constant interior radius solution. In that case, the total radius, which includes the atmospheric thickness (see Sect. \ref{sec:atm_thick}), is outputted, together with an updated total mass that includes the atmospheric mass, $M_{atm}$, obtained as: 

\begin{equation}
    M_{atm} = P_{surf} \frac{4 \pi R_{interior}^{2} }{g_{surf}}
\end{equation}

If $\vert R_{interior, \ i} - R_{guess} \vert$ is greater than the tolerance, the algorithm starts a new iteration, $i+1$, where we update $R_{guess} = R_{interior, \ i}$. The procedure described above is repeated iteratively until the tolerance condition is fulfilled. We let the algorithm run for at least 2 iterations, with a default tolerance of $10^{-3} \times R_{interior, \ i}$. This value can be changed by the user. The planet's internal and equilibrium temperatures, $T_{eq}$ and $T_{int}$, as well as the compositional parameters, CMF and $Z_{env}$, are kept constant across the iterative scheme.

\section{Thermal evolution} \label{sec:thermal_evol}

\subsection{Formalism}
In a simple coupled interior-atmosphere model, the internal temperature, $T_{int}$, can be chosen arbitrarily by the user. However, in the atmospheres and deep interiors of gas giants this parameters depends on the age. For simplicity, in this section we assume that the only source of heat is the secular heat, $L = L_{sec}$, or in other words, $T_{int} = T_{thermal}$. After the planet is formed, it contracts and cools down as it ages because it emits energy, decreasing $T_{int}$. The luminosity to which this emission corresponds is defined as:

\begin{equation} \label{eqn:lumi}
    L = 4 \pi \sigma R_{pl}^{2} T_{int}^4
\end{equation}

\noindent where $R_{pl}$ is the total radius of the planet. The equation that relates the luminosity with time, $t$, is \citep{Thorngren16}:

\begin{equation} \label{eqn:therm_evol}
    \frac{\partial L}{\partial m} = - T \frac{\partial S}{\partial t}
\end{equation}

If Eq. \ref{eqn:therm_evol} is rearranged, and the derivative of the entropy respect to time is isolated, we obtain: 

\begin{equation} \label{eqn:therm_final}
    \frac{\partial S}{\partial t} = - \frac{L}{M} \frac{1}{\int_{0}^{1} dm \ T(m)}
\end{equation}

To integrate the term $\int_{0}^{1} dm \ T(m)$, both the temperature and the enclosed mass profiles are required. These are obtained directly from the interior model.

\subsection{Thermal sequence and solver} \label{sec:sequence}

For a constant planetary mass, composition and equilibrium temperature, we compute a succession of coupled interior-atmosphere models at different $T_{int}$ values. This constitutes a thermal sequence, in which each model represents a thermal state characterized by a set of values for the parameters in Eq. \ref{eqn:therm_final}, $\mathbf{m_{i}}$ = $\{ T_{int, \ i}, L_{i}, S_{top, \ i}, f_{S, \ i} \}$. The luminosity in each state is calculated with the input internal temperature, and the total radius calculated by the interior-atmosphere model, following Eq. \ref{eqn:lumi}. The entropy profile of each interior model, $S(r)$, is evaluated at the outermost boundary (surface), to obtain $S_{top}$. Finally, 
these three parameters together with the $T(r)$ and $m(r)$ interior profiles are used to calculate the function $f_{S, \ i} = f(S_{top})_{i} = \frac{dS}{dt} \big\vert_{i}$ (Eq. \ref{eqn:therm_final}).

The sequence of $i = 1, 2, ..., N$ models is then integrated in time by the solver \verb|scipy.interpolate.odeint|. In addition to the array with the values of $\frac{dS}{dt} \big\vert_{i}$, the solver needs an initial condition for the entropy, $S(t_{0}=0) = S_{0}$. MESA \citep{Paxton11} uses an initial condition for low-irradiation gas giants of $S_{0} = 10.5 \ k_{B} m_{H}$\footnote{$m_{H} = 1.6735 \times 10^{-27}$ kg, and $k_{B} = 1.3806 \times 10^{-23}$ J/K}, which is equivalent to 0.087 MJ/kg/K. In GASTLI, the initial condition can be adjusted by the user, with a default value of $S_{0} = 12 \ k_{B} m_{H}$.


\section{Jupiter} \label{sec:jupiter}

\subsection{Mass-radius relations} \label{sec:Jupiter_mr}

\begin{figure*}[h]
   \centering
   \includegraphics[width=\hsize]{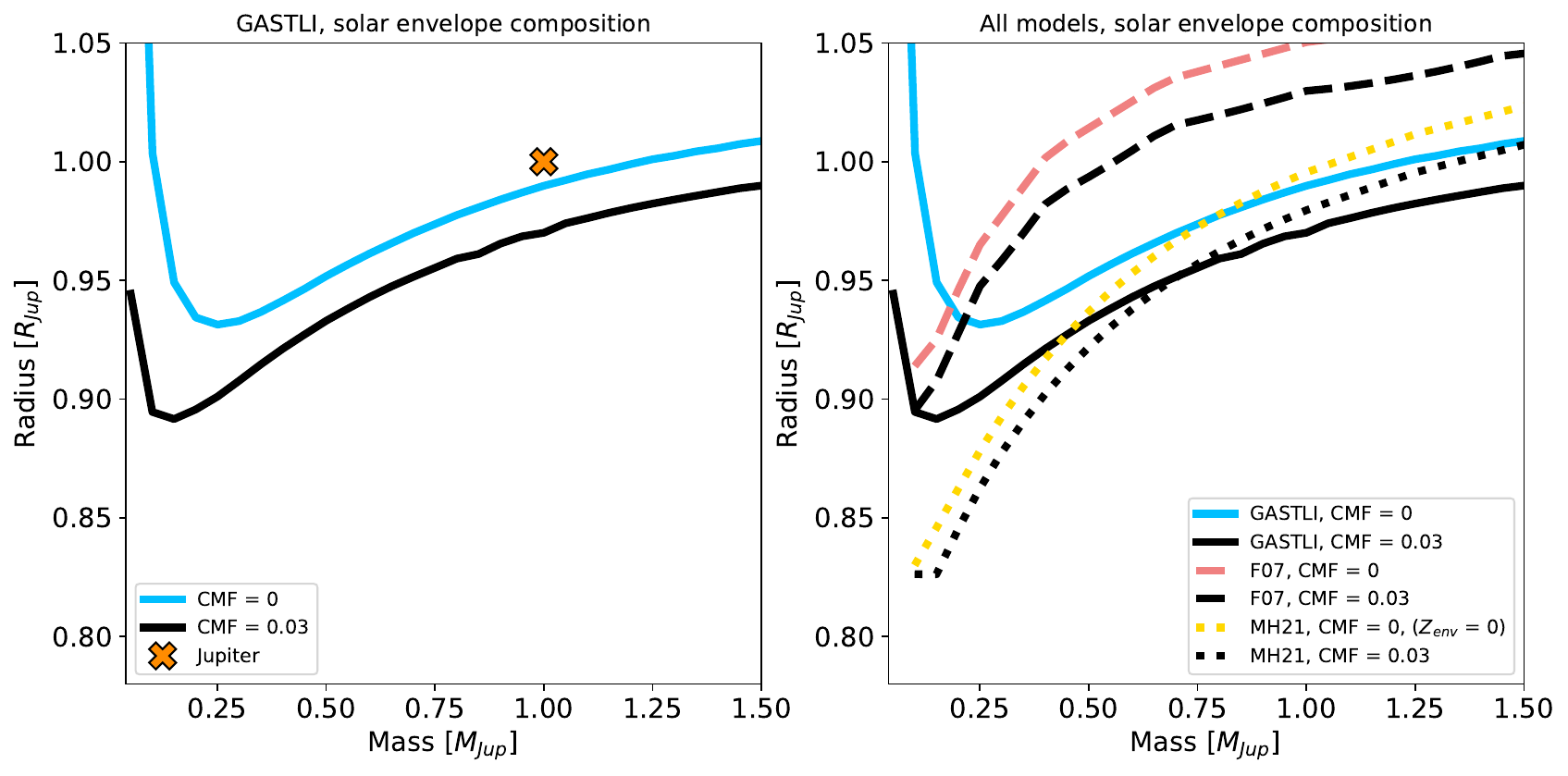}
      \caption{Left: GASTLI models for Jupiter. GASTLI can reproduce Jupiter's mass and radius data with the fiducial model within 1.6\%. Solid lines show the total radius obtained with GASTLI for core mass fractions (CMF) equal to 0 and 0.03 (fiducial) assuming a solar metallicity in the envelope. The global average equilibrium temperature (122 K) and internal temperature (107 K) of Jupiter were adopted .
      Right: Comparison of mass-radius relations between GASTLI and two widely used interior models for gas giants: \cite{Fortney07} (F07) and \cite{Muller21} (MH21). For the same composition, GASTLI agrees within uncertainties with other publicly available mass-radius relations. The fiducial case at CMF = 0.03 is indicated in black for all models. F07 and MH21 models were obtained assuming the irradiation and age of Jupiter. For MH21 at CMF = 0, a solar $Z_{env}$ = 0.013 is not available, so $Z_{env}$ = 0 is assumed instead.}
         \label{fig:Jupiter_MRrel}
\end{figure*}

We start the validation of our model by comparing it to widely used mass-radius relationships for warm gas giants applied to Jupiter. Even though GASTLI can compute thermal evolution tracks (Sect. \ref{sec:thermal_evol}), and thus the evolution of the radius with time, we start by adopting a fiducial internal temperature for Jupiter, $T_{int} $ = 120 K, to eliminate any differences that could rise from assuming a cloud-free atmosphere. We assumed the global average to calculate the equilibrium temperature of Jupiter ($T_{\star}$ = 5777 K, $R_{\star}$ = 1 $R_{\odot}$, $a_{d}$ = 5.2 AU), $T_{eq}$ = 122 K. Jupiter's measured atmospheric metallicity by Galileo and Juno is 2-3 $\times$ solar \citep{Mahaffy00,Li20}, but recent work suggests that its deep atmosphere ($P$ > 30 bar) may have a solar metallicity \citep{Cavalie23,MH24}, which we adopt in our models. Jupiter interior models based on Juno's latest gravitational field measurements constrain its total metal content to 15-20 $M_{\oplus}$ \citep{Howard23}. For a solar envelope composition, at least a CMF = 0.03 is required to have this minimum in metal content of 15 $M_{\oplus}$. Hence, we adopt CMF = 0.03 as our fiducial case. Fig. \ref{fig:Jupiter_MRrel} shows the GASTLI models at three CMF values, including the fiducial case, in comparison to Jupiter's mass and radius. GASTLI's fiducial model reproduces Jupiter's radius within 5\%. For a homogeneous planet with a solar envelope composition (CMF = 0, blue line), GASTLI's model agrees with Jupiter mass and radius data within $\sim$1\%.

\subsubsection{Comparison with other interior models} \label{sec:atm_temp_Jupiter}

We compare our mass-radius relations for Jupiter with models obtained by \cite{Fortney07} (F07). These models include the effect of an upper atmosphere by computing the boundary condition of the interior with a grid of non-grey atmospheric models. The atmospheric composition is assumed to be solar across the grid. Similar to our petitCODE grid, the effect of condensation is taken into account via equilibrium chemistry, but the atmospheric opacity calculations do not consider clouds. For the fiducial case, the F07 model is 0.03 $R_{Jup}$ higher than GASTLI for a Jupiter-mass planet. As a consequence, to match Jupiter's mass and radius, the F07 models need to assume a CMF = 0.075 with a solar envelope composition, which adds up to a total metal mass of 27.7 $M_{\oplus}$. The difference between the F07 models and GASTLI can be explained by the use of a different EOS or by differences in the boundary temperature of the interior. The latter can be caused by the use of different opacity data \citep{Baudino17,Acuna23}. The EOS for H/He used in the F07 models is \cite{Saumon95}, which has been proven to overestimate the radius for a Jupiter analog compared to the \cite{Chabrier19} with non-ideal effects by a few percent \citep{HG23}.

In addition to the F07 models, the right panel in Fig. \ref{fig:Jupiter_MRrel} shows a comparison between GASTLI and models by \cite{Muller21} (MH21). These were obtained with MESA by using a 1:1 rock and water mixture to represent the core and heavy metals in the envelope, as well as MESA's built-in H/He EOS \citep{Chabrier19}. The MH21 models match the mass and radius of Jupiter for a composition of a pure H/He planet (CMF = 0, $Z_{env} = 0$). For the fiducial case, the MH21 model agrees well with GASTLI. However, for planets of similar irradiation and composition, but lower masses, MH21 obtain a radius more than 10\% lower than GASTLI's estimate. The reasons for this may be: 1) MH21 assume a boundary pressure for their interior models significantly higher than the transit pressure, 2) the atmospheric flux assumption in MESA may underestimate the boundary temperature, producing a colder interior, or 3) differences in H-He EOS.


\subsubsection{Effect of atmospheric temperature} \label{sec:atm_temp_Jupiter}

\begin{figure}[h]
   \centering
   \includegraphics[width=\hsize]{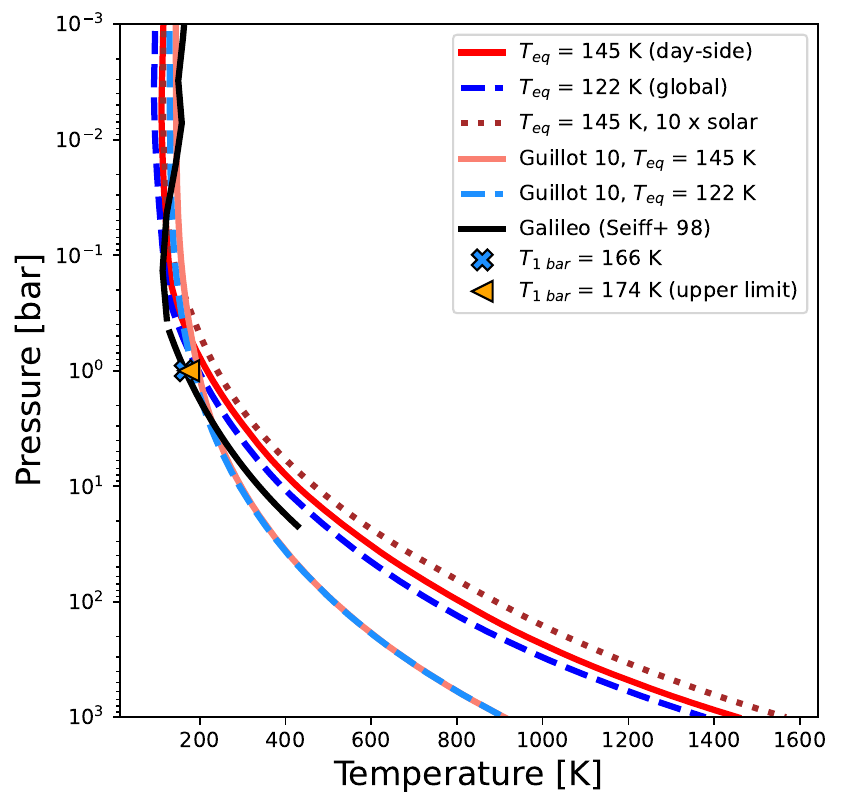}
      \caption{Atmospheric profiles for Jupiter's interior-atmosphere models. Our clear petitCODE fiducial model is 50 K warmer than Jupiter's atmospheric data provided by the Voyager and the Galileo probe \citep{Seiff98,Gupta22,Li24}. We assumed the fiducial case, with an internal temperature of $T_{int} $ = 107 K, a solar composition, and a core mass fraction of CMF = 0.03. For the \cite{Guillot10} atmospheric models, we adopted an infrared opacity $\kappa_{th}$ = 0.01 cm$^{2}$/g, and a visible-to-thermal opacity ratio of $\kappa_{v}/\kappa_{th} = \gamma$ = 0.4.}
         \label{fig:Jupiter_atm_prof}
\end{figure}

We explore the effect of differences in temperature at the bottom of the atmosphere in the total radius. For a Jupiter analog with an internal temperature $T_{int} $ = 107 K, a solar envelope composition, and a CMF = 0.03, we compute interior-atmosphere models to obtain their atmospheric temperature profiles and total radii. The atmospheric profiles are shown in Fig. \ref{fig:Jupiter_atm_prof}, where we consider two models: the day-side (145 K) and the global (122 K) equilibrium temperatures. The temperature at the bottom of the atmosphere differs by 76 K (1456 K and 1380 K, respectively). This difference of 76 K in the day-side average increases the total radius by 0.01 $R_{Jup}$ compared to the global average.

We compare our petitCODE atmospheric model to the analytical semi-grey models by \cite{Guillot10}. For these models, we adopt the recommended opacity values for hot Jupiters, which are a thermal opacity of $\kappa_{th}$ = 0.01 cm$^{2}$/g, and an optical-to-infrared opacity ratio of $\kappa_{v}/\kappa_{th} = \gamma$ = 0.4 \citep{Guillot10}. Analytical semi-gray models are widely used to set boundary conditions in interior and thermal evolution models \citep{Jin14,MacKenzie23,Poser24}, and atmospheric escape models \citep{OW13}. Furthermore, these are implemented in MESA as one of the options to determine the boundary condition \citep{Paxton13}. We show these analytical models for comparison, and do not use them in our bolometric flux estimations at all. Fig. \ref{fig:Jupiter_atm_prof} shows that the grey models evaluated at the opacity values recommended for hot Jupiters can underestimate the bottom temperature by 600 K, which yields a total radius of 0.94 $R_{Jup}$ for both the day-side and global averages. This change in radius is similar to that produced by an enrichment in metals of the envelope from solar to 10 $\times$ solar for an atmospheric temperature that is similar to the global average (dotted line). This highlights the importance of benchmarking grey (or double-grey, in the case of \citealt{Guillot10}) atmospheric models with non-grey models, especially if these are to be used to determine the boundary temperature of interior models. The \cite{Guillot10} models match within 50 K our petitCODE models for a Jupiter analog if 
$\kappa_{th}$ = 0.1 cm$^{2}$/g and $\gamma$ = 0.4  are assumed as opacity values. Hence, we discourage using these values (from hot Jupiter atmospheric models) for colder or smaller planets. \cite{Jin14} provide tabulated values of $\gamma$ for different equilibrium temperatures by benchmarking the analytical \cite{Guillot10} model to numerical atmosphere models.

\subsection{Thermal evolution} \label{sec:jupiter_therm}

\begin{figure}[h]
   \centering
   \includegraphics[width=\hsize]{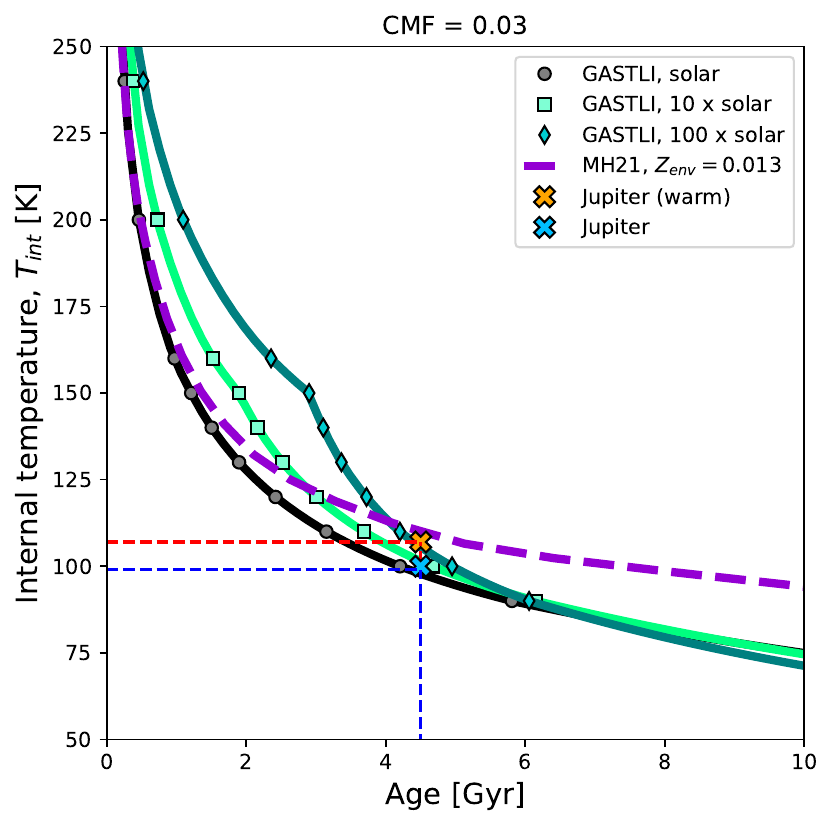}
      \caption{Thermal evolution curves for a Jupiter analog. GASTLI's fiducial thermal curve obtains Jupiter's lower internal temperature estimate at 99 K (cold). An upper estimate of Jupiter's intrinsic temperature (warm, 107 K) is shown for comparison. The fiducial case assumes a solar envelope composition and a CMF = 0.03. GASTLI models at higher envelope metallicities show that cooling takes longer compared to the fiducial case due to more optically thick atmospheres. Markers indicate the internal temperature values at which the entropy was sampled in the thermal sequence (see Sect. \ref{sec:sequence}). A MH21 model for the fiducial case ($Z_{env}$ = 0.013) is shown for comparison.}
         \label{fig:Jupiter_therm_evol}
\end{figure}

We compute the thermal evolution of a Jupiter analog for our fiducial case (CMF = 0.03, log(Fe/H) = 0) assuming the global equilibrium temperature. We also adopt a hot start entropy value of 12 $k_{B} m_{H}$ \citep{Marley07,Spiegel12}. We find that a lower initial entropy (cold start, 9 $k_{B} m_{H}$) does not change the internal temperature at 4.5 Gyr. We can see in Fig. \ref{fig:Jupiter_therm_evol} that GASTLI's fiducial model obtains an internal temperature at the solar system's age (4.5 Gyr) of 97 K. In the past, the effective temperature of Jupiter has been reported as $T_{eff}$ = 125 K \citep{Orton75,Hubbard77,Vazan18}, which requires an internal temperature of $T_{int}$ = 107 K \citep{Li18,apple_model} for a Bond albedo of 0.5. In contrast, Voyager measurements show an intrinsic flux of 5.44 W/m$^{2}$ for Jupiter \citep{Pearl91,MV23}, which corresponds to $T_{int}$ = 99 K. Our fiducial model agrees within 2 K with the latter estimate. 

Nonetheless, the difference between our fiducial model and the warmest observational measurement could be due to our assumption of clear atmospheres with no cloud contribution to the opacities in the atmospheric models. Moreover, thick clouds in the deep troposphere, and superadiabatic gradients can slow down heat release, leading to higher internal temperatures at the same age in comparison to clear atmospheres \citep{Poser19, Poser24,Morley24}. To simulate an increase in atmospheric opacity and see its effect in the thermal evolution with GASTLI, we increase the metallicity in the envelope by $\times$ 10 and $\times$ 100 solar. These two enriched models obtain an internal temperature of 101 and 105 K, respectively. Helium rain and the H-He phase separation can also delay the thermal evolution of Jupiter by releasing energy \citep{Fortney04,apple_model}. 

We also show for comparison the thermal curves by MH21, computed with MESA. These estimate Jupiter's current internal temperature at 110 K, which is closer to the warm estimate of 107 K \citep{Li18,apple_model}. A difference of $\sim$10 K between HG21 and GASTLI could be due to the differences in boundary temperature calculations, as well as in EOS. MH21 use the EOS provided by \cite{Chabrier19}, which do not include non-ideal corrections to the entropy in contrast to \cite{Chabrier21}. These can have non-negligible effects in the entropy, especially at planetary temperatures, which are lower than those found in stars \citep{Chabrier21}.

\section{Saturn and Neptune} \label{sec:Neptune}

We continue the validation of our interior-atmosphere model by generating models for Saturn and Neptune. Saturn has an equilibrium temperature at null Bond albedo of 90 K (see Eq. \ref{eqn:F_Teff} and \ref{eqn:Teff_sum}), whereas Neptune's corresponds to 51 K. The minimum equilibrium temperature in our grid of atmospheric models is 100 K (see Table \ref{tab:atm_grid}). Hence, we approximate their equilibrium temperature to 100 K. The measured intrinsic fluxes for Saturn and Neptune are 2.01 W/m$^{2}$ and 0.43 W/m$^{2}$, respectively \citep{Pearl91,MV23}. These are equivalent to internal temperatures of 77 and 52 K, which we adopt for our fiducial models to generate mass-radius relations for Saturn and Neptune, respectively.

\begin{figure*} 
\centering
\begin{subfigure}[b]{0.49\textwidth}
\centering
\includegraphics[width=\textwidth]{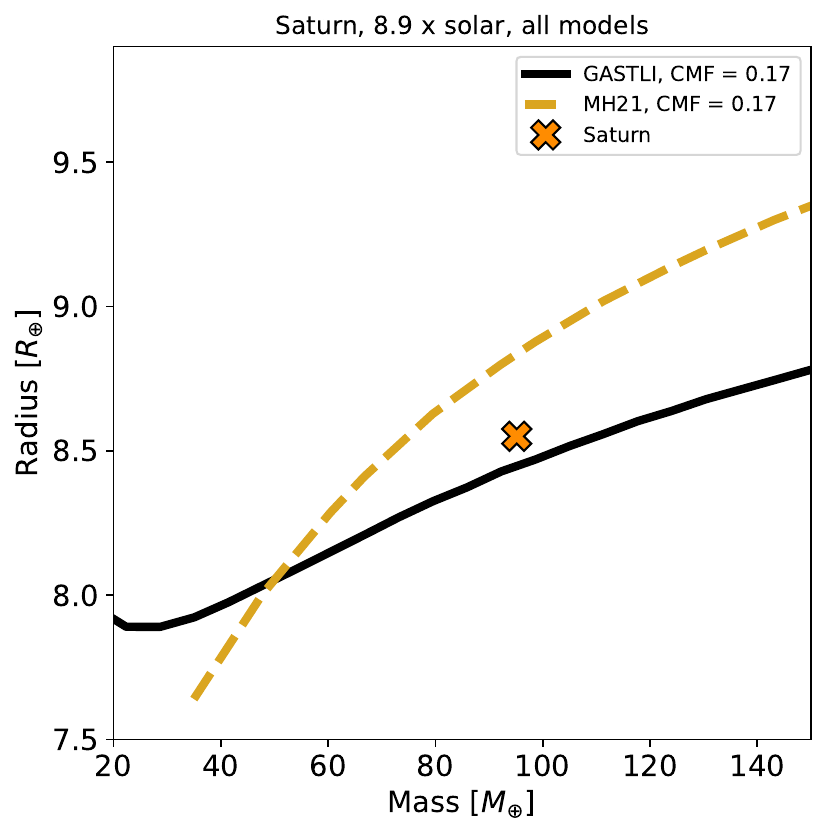}
\end{subfigure}
\hfill
\begin{subfigure}[b]{0.49\textwidth}
\centering
\includegraphics[width=\textwidth]{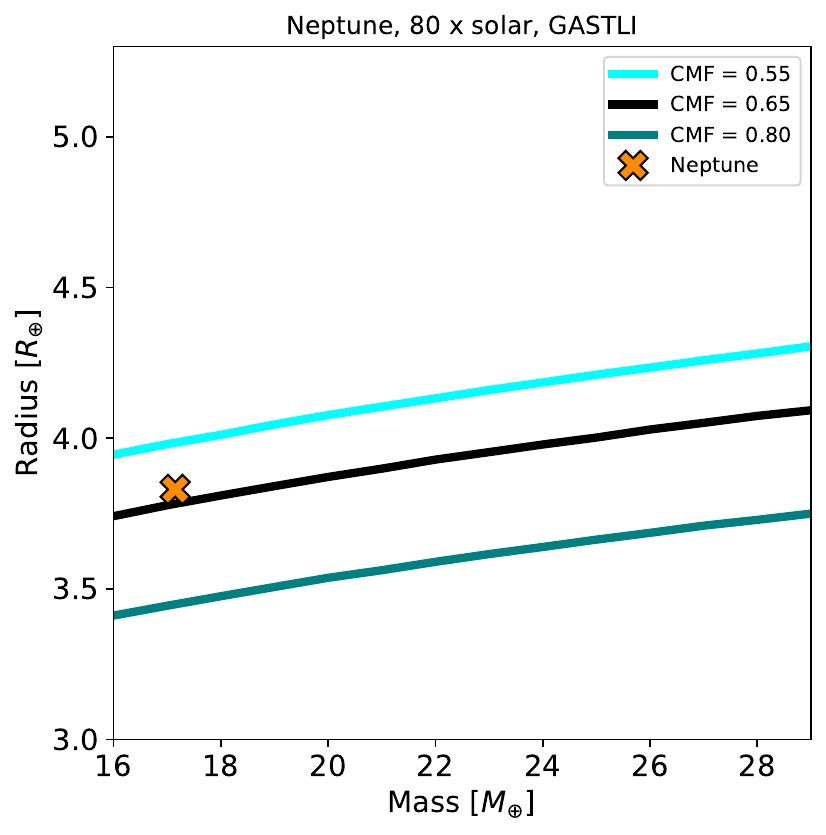}
\end{subfigure}
\caption{Left: Mass-radius relations for Saturn. GASTLI's fiducial model for Saturn can reproduce with a better precision Saturn's mass and radius data than HG21. We assumed an equilibrium temperature of $T_{eq} = $ 100 K across all models, while we adopt an internal temperature $T_{int}$ = 77 for Saturn. Models for Saturn were calculated for a fixed envelope metallicity of 8.9 $\times$ solar, equivalent to $Z_{env}$ = 0.10 for the HG21 model. Right: Mass-radius relations for Neptune. The fiducial model (black) agrees well with mass and radius data. We assumed an internal temperature $T_{int}$ = 52 K for Neptune.}
\label{fig:Saturn_and_Neptune}
\end{figure*}

The atmospheric metallicity of Saturn has been measured as 8.98 $\pm$ 0.34 $\times$ solar \citep{Atreya22}, which we adopt for our fiducial case. In Fig. \ref{fig:Saturn_and_Neptune} (left panel), we show models for Saturn at a constant envelope composition. For MH21 models, this is equivalent to $Z_{env}$ = 0.10. The core mass fraction of our fiducial model is CMF = 0.17. This corresponds to an overall metal mass fraction of $Z_{planet}$ = CMF + (1-CMF) $\times \ Z_{env} = $ 0.25. Models by \cite{Militzer19,MV23} predict a metallic core of 15-18 $M_{\oplus}$ in Saturn, which constrain the maximum metal mass fraction of Saturn at 0.24. For our fiducial model at CMF = 0.17, the total mass of the core is 16.2 $M_{\oplus}$, which is within the uncertainties of these detailed models \citep[see][for a discussion on dilute cores]{Militzer19,Mankovich21}. Saturn's reference radius is the polar radius at 1 bar, which is $R = 8.55 \ R_{\oplus}$, while its mass is $M = 95.16 \ M_{\oplus}$ \citep{MV23}. We modify our coupled model to obtain the total radius at 1 bar instead of 20 mbar for Saturn and Neptune. We can see that GASTLI can reproduce Saturn's mass and data for the fiducial case within 0.10 $R_{\oplus}$, which corresponds to 1\% of its total radius. For comparison, we also compute HG21 models for Saturn's irradiation and an age of 4.5 Gyr. We keep the envelope metal mass fraction constant to $Z_{env}$ = 0.10. The MH21 model for the fiducial case obtains a radius 0.25 $R_{\oplus}$ larger than Saturn's measured radius, which is an error of 2.9\%. To reproduce Saturn's mass and radius, the HG21 models need a CMF of $\sim$ 0.22. This CMF entails a total an overall metal mass fraction of $Z_{planet}$ = 0.30, which is 0.05-0.06 higher than the estimates by our model and the dilute core models.

In contrast to the gas giants, the ice giants of the Solar System have more enriched envelopes. For Neptune, the measured atmospheric metallicity is 80 $\pm$ 20 $\times$ solar \citep{Karkoschka11,Irwin19,Irwin21}. Thus, in Fig. \ref{fig:Saturn_and_Neptune} (right panel) we adopt 80 $\times$ solar for Neptune's fiducial case. This corresponds to log(Fe/H) = 1.90, $Z_{env}$ = 0.57. Similarly to Saturn, the reference radius is the polar radius at 1 bar, $R = 3.83 \ R_{\oplus}$. The reference mass is $M = 17.15 \ M_{\oplus}$ \citep{MV23}. Detailed models of Neptune that take into account dilute cores and gravitational data obtain a range of 0.8-0.9 for the total metal mass fraction \citep{Helled11,Podolak19}. For our fiducial case with an envelope of 80 $\times$ solar composition, a total metal mass fraction of 0.85 corresponds to CMF = 0.65, which is the black line in Fig. \ref{fig:Saturn_and_Neptune} (right panel). GASTLI's fiducial model for Neptune agrees with the data within 0.07 $R_{\oplus}$, which consists of 1.8\% of its total radius. For reference, we also provide the mass-radius relations of the lower and upper limits of Neptune's total metal mass fraction.

\section{HAT-P-26 b} \label{sec:hat_p_26b}

In this section, we showcase the applicability of GASTLI to model warm exoplanets with higher equilibrium temperatures than the Solar System gas and ice giants. We will first explore the effect of the envelope metallicity in the estimation of bulk metallicity and thermal evolution for the well-characterised planet HAT-P-26 b (Sect. \ref{sec:hat_p_26b_forward}). HAT-P-26 b has a mass similar to Neptune ($M = 0.059\pm0.007 \ M_{Jup} = 18.76\pm2.23 \ M_{\oplus}$), but is significantly more irradiated, with an equilibrium temperature of $T_{eq} = 1001$ K \citep{Hartmann11,Athano23}. Its atmosphere was characterised by transmission spectroscopy with the Hubble Space Telescope (HST), revealing a clear atmosphere with an atmospheric metallicity of Fe/H = 4.8$_{-4.0}^{+21.5} \ \times$ solar \citep{Wakeford17}, which is significantly less enriched than Neptune. \cite{MacDonald19} re-analysed HST and Spitzer data \citep{Stevenson16}, obtaining a slightly higher estimate for the atmospheric metallicity of Fe/H = 18.1$^{+25.9}_{-11.3} \ \times$ solar. Nonetheless, a recent atmospheric retrieval analysis by \cite{Athano23} that includes optical data found a lower estimate for the water abundance, hence lowering the atmospheric metallicity. Thus, in this section we adopt the conservative estimate in atmospheric metallicity by \cite{Wakeford17}.

Besides well-determined mass, radius ($R = 0.565^{+0.072}_{-0.032} \ R_{Jup} = 6.33^{+0.81}_{-0.358} \ R_{\oplus}$), and atmospheric metallicity, the age of the host star is characterized as 9.0$^{+3.0}_{-4.9}$ Gyr \citep{Hartmann11}. Hence, all parameters necessary to characterise its total metal content (both envelope and core mass) are available. In Sect. \ref{sec:hat_p_26b_retrieval}, we revisit the metal mass fraction estimate of HAT-P-26 b by performing a Markov chain Monte Carlo (MCMC) retrieval.

\subsection{Effect of CMF and atmospheric metallicity} \label{sec:hat_p_26b_forward}

\begin{figure*} 
\centering
\begin{subfigure}[b]{0.49\textwidth}
\centering
\includegraphics[width=\textwidth]{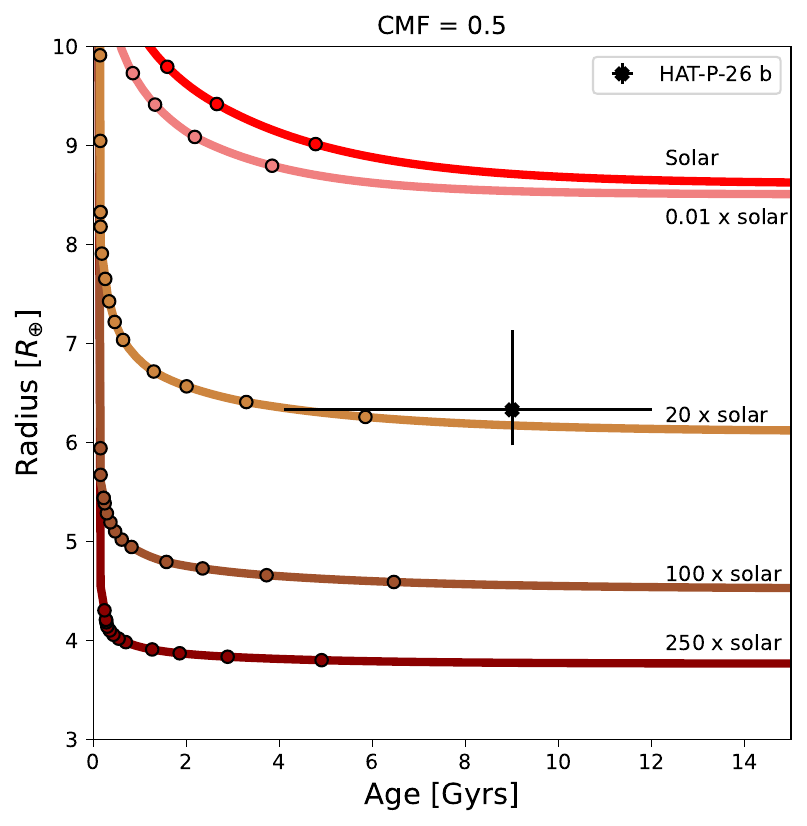}
\end{subfigure}
\hfill
\begin{subfigure}[b]{0.49\textwidth}
\centering
\includegraphics[width=\textwidth]{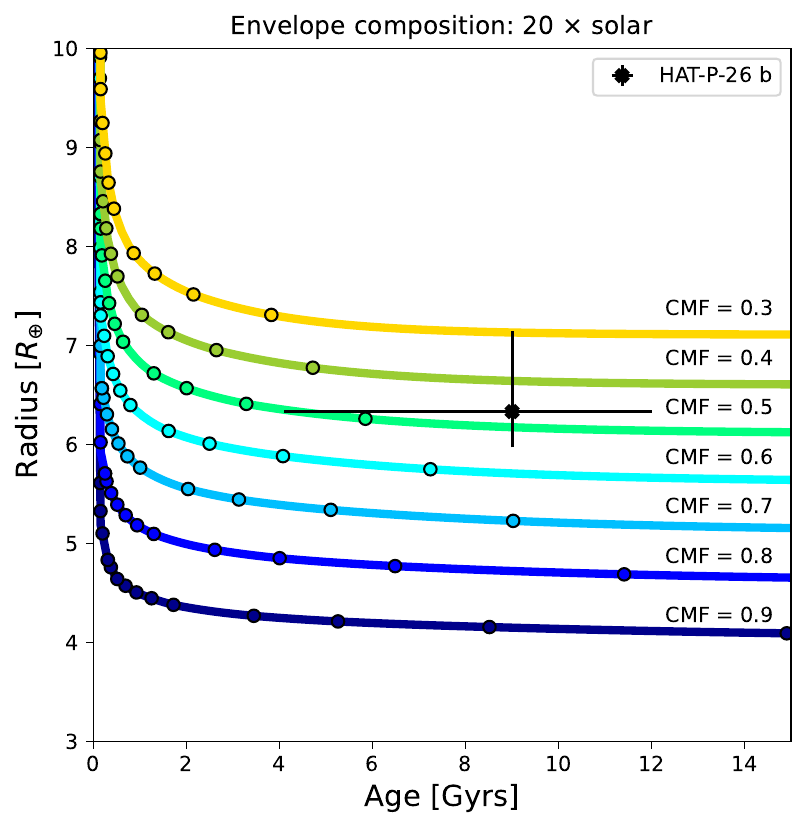}
\end{subfigure}
\caption{Left: Radius evolution as a function of age for HAT-P-26 b for a constant CMF = 0.5. Right: Radius evolution as a function of age for HAT-P-26 b for an envelope metallicity of 20 $\times$ solar. Radius decreases with increasing core mass fraction and age. We adopt HAT-P-26 b's mean mass ($M = 18.76 \ M_{\oplus}$) and equilibrium temperature $T_{eq} \sim$ 1000 K across all models.}
\label{fig:hatp26b_radius}
\end{figure*}

Fig. \ref{fig:hatp26b_radius} shows HAT-P-26 b's radius-age diagram for a constant CMF and varying envelope metallicities (left), and vice versa (right). In both panels, we can see that the radius decreases with age for a constant composition, as expected. From the right panel alone, we observe that the radius decreases with increasing CMF for the same age and envelope composition. Similarly, the left panel shows that for a constant CMF, metal-rich envelopes reduce the density of the planet. The sub-solar metallicity thermal curve is an exception since it is more dense than the solar case. This is due to a warmer deep atmosphere in the solar case, despite the sub-solar case having an almost pure H/He composition ($Z_{env} = 10^{-4}$), in contrast to the solar metal mass fraction in the envelope, $Z_{env} = 0.013$. For models whose envelope is more metal rich, the decrease in radius is particularly noticeable from solar to 20 $\times$ solar, and from 20 to 100 $\times$ solar. These changes are equivalent to envelope metal mass fraction changes from $Z_{env} =$ 0.013 to $Z_{env} = $ 0.26, and from 0.26 to $Z_{env} =$ 0.60, respectively. The effect of the density of the metals in the envelope is enough to produce a decrease in radius of 1-2 $R_{\oplus}$ for a Neptune-mass planet, which is up to 30\% of the total radius. 

The density (and hence the radius) of a planet is not just a result of composition, mass and irradiation, but also of age. Thus, younger planets are less dense than older planets under the same composition, mass and irradiation \citep{Burrows97,Baraffe2003,Baraffe08,LF14,Vazan15,Vazan18,Linder18,Muller21}. In Fig. \ref{fig:hatp26b_radius}, the difference in radius for the same mass and composition between 1 and 4 Gyr is 0.5 $R_{\oplus}$, whereas between 4 and 9 Gyr this corresponds to 0.1 $R_{\oplus}$. Despite having a longer interval of time in the latter, the difference is greater at younger ages due to the very rapid contraction for ages under 3 Gyr. This is the result of the fast decrease of internal temperature: within the first Gyr of age, a planet internal temperature can vary several hundreds of K. This  can be explained by the Kelvin-Helmholtz timescale, $\tau_{KH}$, being significantly shorter at young ages, since the cooling luminosity is proportional to the luminosity, $L$ (Eq. \ref{eqn:lumi}), and $\tau_{KH} \propto 1/L$. The Kelvin-Helmholtz timescale is also dependent on radius, although after the early contraction phase the radius is approximately constant.

The difference in radius between different ages decreases with higher metal content: for a 250 $\times$ solar envelope and CMF = 0.5, the 1 Gyr-old radius is very similar to the radius at 4 Gyr, while for the solar envelope, the radius decreases 1 $R_{\oplus}$ between 1 and 4 Gyr, which is 16\% of the total radius. Hence, this highlights the importance of determining the stellar age precisely, especially for younger planets, where there is a strong degeneracy between age and bulk metal content \citep{Muller22}.

\subsection{Interior composition retrieval} \label{sec:hat_p_26b_retrieval}
In this section we revisit the interior composition of HAT-P-26 b with an interior retrieval to estimate its bulk metal content, which can be compared with predictions from core accretion or other formation mechanisms.

We generate a grid of GASTLI models for fast interpolation in the retrieval. The grid spans masses from 15 to 30 $M_{\oplus}$ with step size $\Delta M$ = 0.5 $M_{\oplus}$, CMF = 0 to 0.99 with $\Delta$CMF = 0.1, log(Fe/H) = -2 to 2 with $\Delta$log(Fe/H) = 1, and $T_{int}$ = 50 to 350 K with $\Delta T_{int}$ = 60 K. The final grid consists of 6000 interior structure models in total. For internal temperatures below 50 K, the solution function provided by the ODE solver (see Sect. \ref{sec:sequence}) is used to calculate their corresponding ages. We used the emcee package \citep{emcee} as Markov chain Monte Carlo (MCMC) sampler. The priors are uniform for CMF = $\mathcal{U}(0.1,0.99)$, and $T_{int}$ = $\mathcal{U}(0,100)$. The mass and log(Fe/H) priors are Gaussian distributions, with the respective mean and uncertainties of the data. The likelihood is calculated following \cite{Dorn15,Acuna21} (their equations 6 and 14, respectively). In the case of HAT-P-26 b, we use the mass, radius, age and atmospheric metallicity as observable parameters to calculate the likelihood. With emcee, we run 32 walkers and N = $10^{5}$ steps. We computed the autocorrelation time for our retrieval, $\tau \sim$ 100, so $10^{5}$ steps is a sufficiently long chain to ensure convergence with $\tau << N/50$.

\begin{figure*}[h]
   \centering
   \includegraphics[width=\hsize]{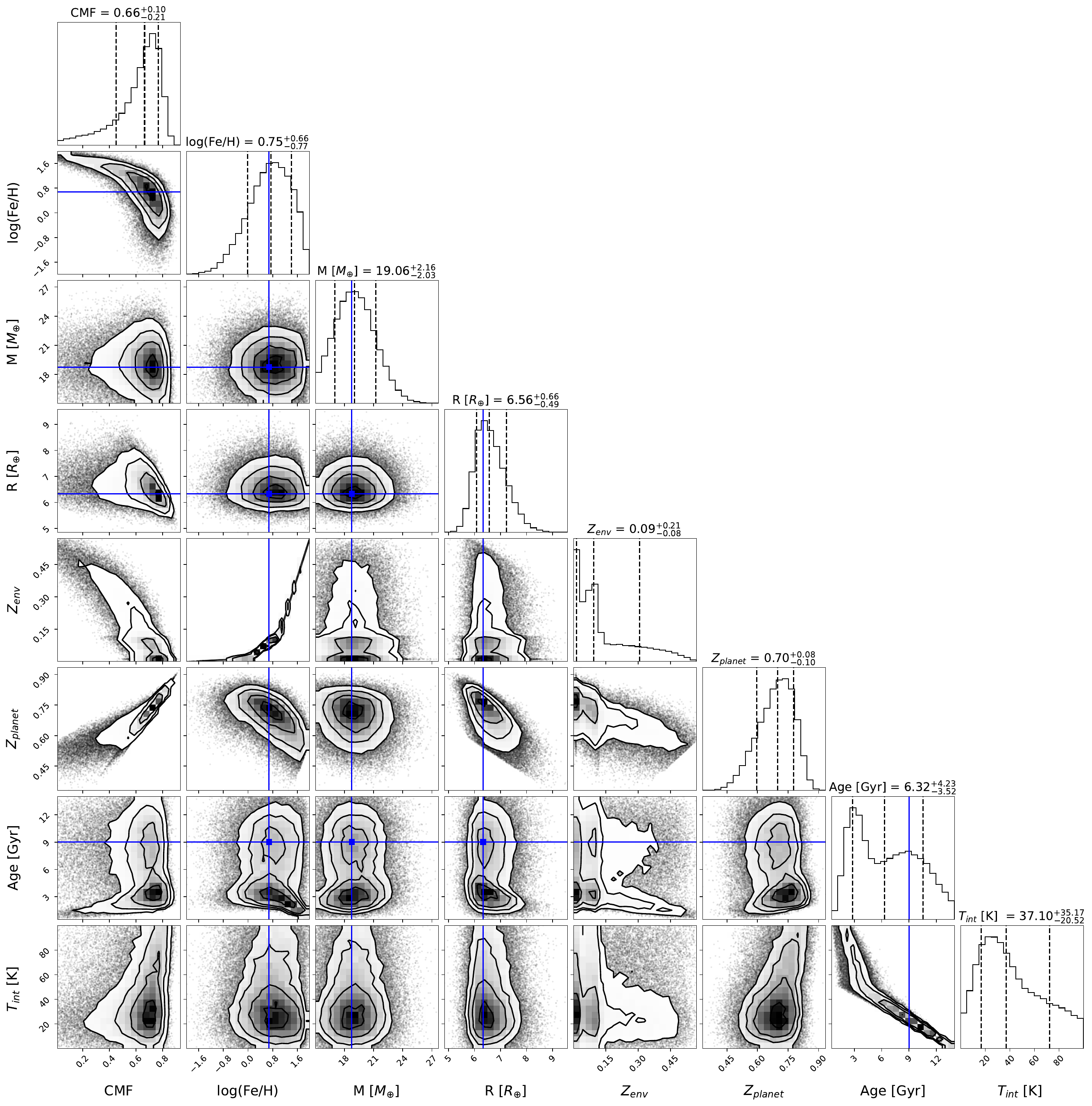}
      \caption{2D and 1D marginalised posterior distribution functions for our MCMC retrieval on HAT-P-26 b data. Blue lines indicate the mean value of the observed data for the mass, radius, atmospheric metallicity and age. Black lines in the 2D distributions indicate the 1, 2 and 3 $\sigma$ regions.}
         \label{fig:hatp26b_corner}
\end{figure*}

Fig. \ref{fig:hatp26b_corner} shows the corner plot corresponding to our interior retrieval for HAT-P-26 b. The posterior distributions of the observables parameters (mass, radius, atmospheric metallicity and age) agree well with the observed values (blue). The retrieved total metal mass fraction of HAT-P-26 b is constrained as $Z_{planet}$ = $0.70^{+0.09}_{-0.10}$, which corresponds to a CMF = $0.66^{+0.10}_{-0.21}$. We compute the total planet bulk metal mass fraction, $Z_{planet} =$ CMF + (1-CMF) $\times \ Z_{env}$. Hence, the metal bulk content of HAT-P-26 b ranges from $Z_{planet} = $ 0.60 to 0.78 at 1-$\sigma$ confidence. \cite{Hartmann11} estimate the possible range of this parameter by comparing mass and radius with mass-radius relationships with interior models by \cite{Baraffe08}, constraining $Z_{planet}$ between 0.5 and 0.9. The combination of our interior-atmosphere models with the measured atmospheric metallicity reduces the degeneracy in bulk metal mass fraction, from 0.9 to 0.78 for the 1-$\sigma$ upper limit, and from 0.5 to 0.6 in the 1-$\sigma$ lower limit.

\begin{figure}[h]
   \centering
   \includegraphics[width=\hsize]{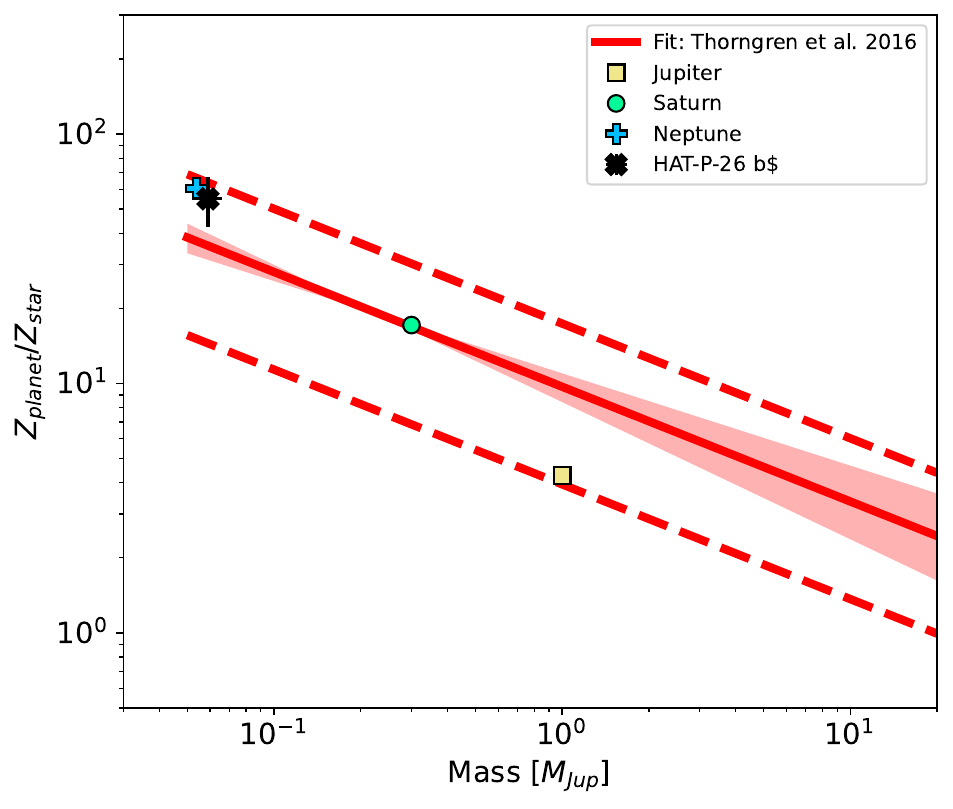}
      \caption{Mass-bulk metallicity diagram. HAT-P-26 b metal content ratio is compatible with the exoplanet population within the 1$\sigma$ posterior predictive \citep[dashed line,][]{Thorngren16}, and Neptune \citep{Podolak19}. The mass-bulk metallicity fit is adopted from \cite{Thorngren16}, (9.7$\pm$1.28) $M_{planet}^{-0.46\pm0.09}$. Estimates for Jupiter and Saturn are shown for comparison \citep{Howard23,Militzer19}.}
         \label{fig:mass_met_hatp26b}
\end{figure}

We compare our bulk metal mass fraction to the metal content of the star to determine the mass-metallicity relation for HAT-P-26 b. We calculate the metal content of the star as $Z_{star} = 0.014 \times \ 10^{[Fe/H]}$, following \cite{Thorngren16,Teske19}. We adopt the metallicity of the star from \cite{Hartmann11}, [Fe/H] = -0.04$\pm$0.08. We show the position of our current $Z_{planet}/Z_{star}$ estimate for HAT-P-26 b in the mass-bulk metallicity diagram in Fig. \ref{fig:mass_met_hatp26b}. The error bars of this estimate is obtained by sampling Gaussian distributions defined by the mean and uncertainties of our interior retrieval $Z_{planet}$ estimate and of the stellar host metallicity. Our estimate is compatible with the fit within 1$\sigma$. Previous estimates in the bulk metal content with an upper limit of $Z_{planet} = $ 0.9 \citep{Hartmann11} produce a $Z_{planet}/Z_{star}$ ratio outside the 1$\sigma$ predictive of the fit, initially suggesting that HAT-P-26 b was more enriched in metals than the rest of the gas giant population. Moreover, HAT-P-26 b bulk metal mass fraction is lower ($Z_{planet}$ = 0.60-0.78) than that of Neptune ($Z_{planet}$ = 0.80-0.90), but the $Z_{planet}/Z_{star}$ ratio of both planets agree within uncertainties. This could be due to a lower metallicity of the host star compared to the Sun, although its metallicity is compatible with the solar value within uncertainties. A refinement of the host star [Fe/H] would be required to confirm this. A metal-poor star, together with weak core erosion, could contribute to explain why HAT-P-26 b's atmospheric metallicity is lower than Neptune's. If the atmospheric metallicity of HAT-P-26 b's is refined and is in the higher end value (26.3 $\times$ solar), as little as 45\% of the planet mass (8.5 $M_{\oplus}$) could be locked in the core. This would be slightly lower than what formation models have assumed for HAT-P-26 b so far \citep{Alidib18}.

\section{Warm extrasolar gas giants} \label{sec:exoplanets}

\begin{figure*}[h]
   \centering
   \includegraphics[width=\hsize]{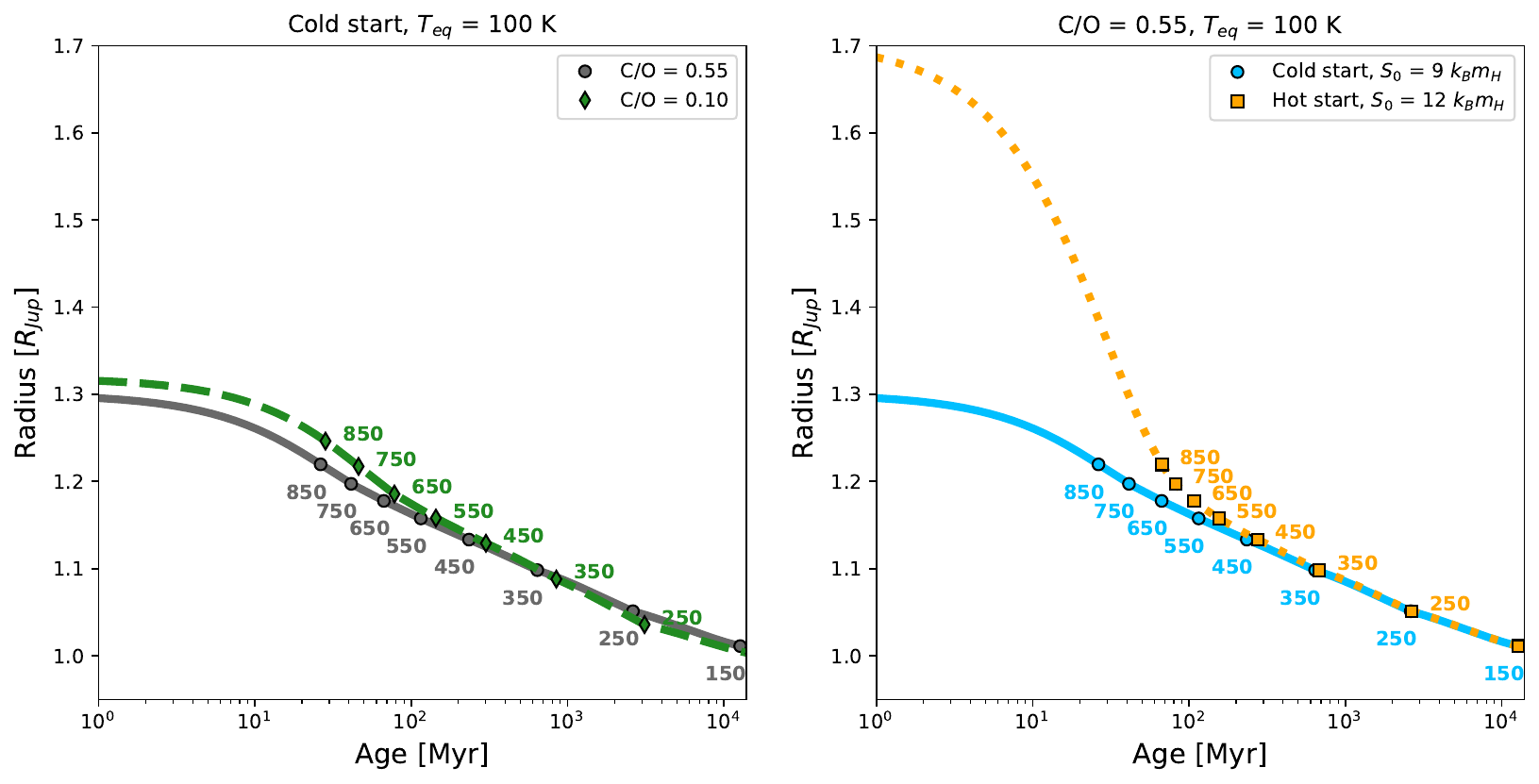}
      \caption{Left: Radius evolution of a 5 $M_{Jup}$ assuming different C/O ratios. A low C/O ratio can slightly increase the planet radius by 2\% at young ages (< 100 Myr). Right: Radius evolution of the same planet at a constant C/O ratio and different initial entropy values. A hot start can increase the planet radius significantly for ages < 100 Myr. We adopt an equilibrium temperature of 100 K across all models. Markers are labelled with their corresponding internal temperature, $T_{int}$.}
         \label{fig:massive_thermal_evol}
\end{figure*}

In this section, we show the effect of C/O and initial entropy on the thermal evolution of massive warm gas giants. Our petitCODE grid includes atmospheric models at C/O = 0.55 and 0.10. In Fig. \ref{fig:massive_thermal_evol}, we explore the effect of the C/O ratio in the thermal evolution of a 5 $M_{Jup}$ planet at a low equilibrium temperature, $T_{eq}$ = 100 K. In the left panel, we can see that the low C/O model presents a radius 2\% larger than the solar value models, which is produced by warmer temperatures at the bottom of the atmosphere. We can see that this effect applies to planets younger than 100 Myr. At that point in their thermal evolution, planets at different C/O ratios present similar radii. If we compare their internal temperatures, a lower C/O ratio delays the thermal cooling of the envelope slightly, by a difference of $\sim$ 100 Myr.

The initial entropy with which a planet forms depends on the formation model and their assumed parameters, such as core accretion rates. The default initial entropy in GASTLI is 12 $k_{B}m_{H}$, which corresponds to a hot start in planet formation models. This is associated with planets formed via disk instability, in contrast to a cold start in core accretion models. \cite{Marley07,Spiegel12} show that the initial entropy for a 5 $M_{Jup}$ can range from 9 $k_{B}m_{H}$ to 12 $k_{B}m_{H}$. Thus, we compare the effect of these two values of the initial entropy for a constant envelope composition in Fig. \ref{fig:massive_thermal_evol}. In the hot start case, the thermal cooling of the planet is delayed by $\sim$ 80 Myr, until 200 Myr in age, where the hot and cold start models are indistinguishable. These trends are similar for all equilibrium temperatures ($T_{eq}$ < 1000 K, not shown).

\section{Discussion} \label{sec:discussion}

In this section, we summarize and discuss the effects of several assumptions that differ between interior models that have an effect in radius and thermal evolution calculations, such as EOS, cloudy atmospheric models, dilute cores, and the top boundary pressure. 

\subsection{Sensitivity to EOS}

In this work, we validated our interior-structure model by comparing it to other widely used interior models. The final radius and density obtained by each model is highly dependent on the EOS used. For example, in Sect. \ref{sec:Jupiter_mr}, we found that interior models that employ the SCvH95 EOS for H/He, such as \cite{LF14,Thorngren16}, overestimate the radius for a Jupiter analog by 5\% compared to the CD21 EOS. This is in agreement with the comparison carried by \cite{HG23}. \cite{HG23} show that for a constant pressure, the density can be underestimated per unit pressure in a pure H/He interior model by \cite{Saumon95} compared to the other EOS for $P>$ 10 GPa (their figure 4). A pure H/He composition can therefore lead to an overestimation of the amount of metals when trying to reproduce the mass and radius with a forward interior model, or performing interior structure retrievals. A difference of 0.05 $R_{Jup}$ in a Jupiter analog entails an enrichment of 16 $M_{\oplus}$ in the core (5\% of the planet's total mass). These differences in CMF are similar to the uncertainties obtained from retrievals for gas giants for the bulk metal mass fraction with current mass, radius and age uncertainties \citep{Muller22}. Consequently, the use of outdated EOS in interior model retrievals may result in the shift of the posterior distribution function by at least one standard deviation, especially if the radius and age are known at high precision. Neptune-mass planets such as HAT-P-26 b, and sub-Neptunes with a non-negligible amount of H/He in their envelope \citep[see][for metal content in the envelope of K2-18 b and TOI-270 b, respectively]{Madhu20,Benneke24}, may be more sensitive to the choice of EOS since the change in radius is larger relative to their small size \citep{haldemann24}. 


\subsection{Sensitivity to atmospheric opacity}

Clouds and metal enhancement both slow the cooling of the planetary interior. In Sect. \ref{sec:jupiter_therm}, we demonstrated that an enhancement in atmospheric metallicity from solar to $ \times$100 solar in a clear model can produce an increase in Jupiter's internal temperature of 10 K. Similarly, \cite{Linder18} find that their fiducial cloud solar models and clear solar models have a similar difference to that seen between the solar and the 4 $\times$ solar clear models. For their cloudy models, they assume a cloud settling parameter of $f_{sed} = 0.5$. The cloud settling parameter is the ratio of the mass averaged grain settling velocity and the atmospheric mixing velocity. For giant exoplanets, cloud models typically adopt $f_{sed}$ values between 0.3 and 3 \citep{Molliere17}. For Jupiter's equilibrium temperature, this parameter is too low, producing high-altitude clouds instead of deep, low-altitude clouds. This discrepancy may explain why we need to enhance the metallicity to more than 4 $\times$ solar to mimic the effect of Jupiter's deep clouds in the thermal evolution. This is further supported by \cite{Poser19}, who show that high clouds have a negligible effect on the thermal evolution of gas giants, whereas deep clouds cause warmer interiors. In general, for the same age and mass, applying an interior model that assumes clear atmospheres to a planet with confirmed deep clouds will underestimate the bulk metal mass fraction, since less metals are needed to match the ($M,R$) data for a colder interior. This is why we limited our analysis to HAT-P-26 b, whose NIR transmission spectrum does not require clouds to account for the opacity sources \citep{MacDonald19}. Upcoming ground-based high-resolution observations of HAT-P-26 b will help constrain the extent of cloud coverage \citep{Panwar22}. In the latter scenario, our bulk metallicity estimates would be a lower limit in the amount of metals in HAT-P-26 b. Furthermore, GASTLI has been developed in a modular setup so that users can couple their own atmospheric PT and metal mass fraction profiles to the interior model by inputting an atmospheric grid file of the same format as our default grid. This will allow for a straightforward comparison of radius and thermal evolution calculations between different atmospheric models, including different cloud models, opacity tables and line data, chemical disequilibrium models \citep{Hubeny07,Visscher11,Zahnle14,Marley15,vulcan_21,Fortney20,Ohno23}, and 3D global circulation models \citep[GCM,][]{Showman02,Showman11,SMartinez19,thor_20,Komacek22,Zhang23}.

\subsection{Sensitivity to core diluteness}

Our models assume that the envelope has a homogeneous distribution of metals, i.e the mass fraction of metals is constant with radius, $Z(r) = Z_{env}$. In addition, the envelope is fully isentropic (adiabatic and reversible), and in the interface between the core and the envelope, the metal mass fraction has a sharp decrease. This compact core plus homogeneous envelope structure is different from the dilute core structure that gravity and ring seismology data suggest for Jupiter and Saturn \citep{Wahl17,Mankovich21,Miguel22}. The dilute core would consist of a region between the compact core (if any) and the observable envelope where metal elements are well-mixed with H/He at higher concentrations that the outer envelope. Dilute cores are easily explained by planet formation mechanisms, such as convective mixing and planetesimal/pebble settling \citep{Lozovsky17}. If a homogeneous and adiabatic interior model is applied to a planet with a highly dissolved core, the bulk metal mass fraction would be underestimated. This is due to hotter non-adiabatic temperature profiles formed as a result of different heat transport regimes, which are determined by the Ledoux and Schwarzschild criteria \citep{Leconte12,Vazan18,apple_model}. Similar to the effect of clouds, a warmer interior is less dense for a fixed mass and composition, so more metals are needed to reproduce similar $(M,R)$ data than an adiabatic interior. Homogeneous models obtain a total metal mass of 15-20 $M_{\oplus}$ for Jupiter, while self-consistent dilute core models calculate a dilute core mass of $\sim 60 \ M_{\oplus}$ for Saturn, and $\sim 100 \ M_{\oplus}$ for Jupiter \citep{Helled17,Helled23,Helled24_rev}. This compact core assumption also impacts thermal evolution, as dilute cores produce higher internal temperatures and luminosities for a fixed age \citep[][and references therein]{Helled24_rev}. Hence, we limited the validation of our model for Jupiter and Saturn to previous estimates from adiabatic and homogeneous interior models.

The temperature profile can also deviate from a homogeneous, adiabatic profile when helium separates from liquid metallic hydrogen and helium rain occurs. \cite{Fortney04} find that helium rain can increase the luminosity of a planet at a constant mass and age, compared to a homogeneous model. This is the case of Saturn, for which \cite{Fortney_Hubbard2003,apple_model} model the evolution of inhomogeneous interiors due to helium phase separation and compositional gradients by adding a term dependent on the derivative dP/d$\rho$ to Eq. \ref{eqn:therm_evol} (their equation 20 and 3, respectively). \cite{Fortney_Hubbard2003} show that inhomogeneity delays the thermal cooling of the interior, in other words, Saturn's effective temperature takes longer to reach its current value in non-homogeneous models. We do not include thermal models for Saturn in Sect. \ref{sec:Neptune} because non-homogeneous interior models are beyond the scope of our paper, but it will be the implemented in future work. In contrast, helium rain is not expected to affect exoplanets with effective temperatures above 140 K as these planets cool down to their equilibrium temperature before the onset of helium separation \citep{Fortney04}. As future missions will increase the semi-major axis at which giant planets are detected, such as PLATO \citep{plato_ref}, the inclusion of helium rain will be required to interpret mass, radius and age data with thermal evolution models. This will be the scope of future work with GASTLI.

For extrasolar gas giants, it is not possible to distinguish between a compact and a fuzzy core from mass, radius, age and atmospheric metallicity alone, given the degeneracy between the mass and extent of the core in radius \citep{Bloot23}. Consequently, new observables such as Love numbers \citep{Love11} are needed to break these degeneracies \citep{Kramm11,Baumeister23}. Love numbers \citep{Kramm11}, and the determination of the internal temperature via disequilibrium chemistry \citep[see][for the case of WASP-107 b]{Sing24,Welbanks24}, can help constrain the temperature profile in the deep envelope. For the vast majority of exoplanets, which only have mass, radius, age and atmospheric metallicity measured, the parameter space is too degenerate to determine whether the temperature profile follows an adiabatic profile or not. To provide an open-source model to interpret these upcoming data, our future work will focus on implementing non-adiabatic self-consistent temperature profiles for the deep interior in GASTLI.

Love numbers have only been measured for hot Jupiters that are tidally deformed \citep{Buhler16,Hardy17,Csizmadia19,Hellard20,Barros22,Bernabo24}. Nonetheless, JWST \citep{Hellard20,Akinsanmi20} and Ariel \citep{ariel_ref} may have enough the precision and cadence in transit photometry to detect oblateness and determine the Love number of warm gas giants with $T_{eq}$ < 1000 K \citep{Akinsanmi20,Berardo22}. For these planets, GASTLI would be a useful tool to provide the density profiles required to compute Love numbers to interpret the data.


\subsection{Sensitivity to transit pressure}

Lastly, we also discussed how the assumed transit pressure has an effect in the total planet radius. The radius for Solar System planets is obtained from radio-occultation experiments \citep{MV23}, which probe the radius at an atmospheric pressure of 1 bar. Hence, this is not an issue for interior models of Jupiter, Saturn and Neptune. However, the radius of exoplanets is determined by transit photometry, which probe the radius at $\sim 20$ mbar \citep{Grimm18}. Thus, interior models that do not include the contribution of the atmosphere from 1 bar to 20 mbar for historical reasons may underestimate the radius for a fixed mass and composition. For a Jupiter analog, the thickness from 1000 bar to 1 bar is twice the thickness from 1 bar to 20 mbar approximately, making a very small difference due to its high surface gravity. However, for a Neptune mass planet, the atmospheric thickness is $\sim 1.5 \ R_{\oplus}$. As we discussed for Jupiter, the pressures between 1 bar to 20 mbar constitute 1/3 of the atmospheric thickness, which would correspond to 0.45 $R_{\oplus}$ for a Neptune-mass planet. This is 10\% of the planet's total radius, for a Neptune analog. Consequently, even though for a Jupiter-mass planet the inclusion of the upper atmospheric thickness may be a small difference (< 3\% in radius), for a Neptune-mass planet the difference is comparable to that caused by different EOS.

\section{Conclusions} \label{sec:conclusion}

The composition of the extrasolar gas giant population may be quite varied, and interior models are required to compute their bulk metal contents, which provides valuable constraints on planet formation and evolution. We present the GAS gianT modeL for Interiors (GASTLI), a coupled interior-atmosphere model applicable to exoplanets whose mass ranges from Neptune's ($M \sim 17 \ M_{\oplus}$) to a few times more massive than Jupiter ($M \sim 6 \ M_{Jup}$). GASTLI is an easy-to-use Python package that is publicly available on GitHub, with its documentation and example cases. The interior model is based on previous work \citep{Brugger16,Brugger17,Mousis20,Acuna21,Aguichine21}, and has been modified to use up-to-date equations of state (EOS) for rock, water and H/He \citep{sesame,Miguel22,Mazevet19,Haldemann20,Chabrier21}. Our interior model assumes that convection is the heat transport mechanism, rendering our model applicable to planets with equilibrium temperatures below $T_{eq} < 1000$ K due to the hot Jupiter radius inflation mechanism. The interior is coupled to a grid of cloud-free atmospheric models obtained with petitCODE \citep{Molliere15,Molliere17} to self-consistently determine the atmosphere-interior boundary temperature, $T(P=1000$ bar), and the contribution of the atmosphere to the total radius (up to 20 mbar). The possible envelope compositions span from sub-solar metallicities (0.01 $\times$ solar) to 250 $\times$ solar (equivalent to a metal mass fraction of $Z_{env}\sim$ 80\%). Our framework enables the calculation of mass-radius relations for a fixed internal temperature (or similarly, luminosity), as well the evolution of the radius and internal temperature with planetary age. 

We succesfully validate our model by generating models for the Solar System gas giants and Neptune. Our estimates in bulk metal content, envelope metallicity and core mass fraction are in agreement within uncertainties with the mass, radius and internal temperature data as well as previous homogeneous models \citep[see][]{Militzer19,MV23,Helled11,Podolak19}. 

As a first showcase of the applicability of our model to exoplanets, we infer the bulk metal content of the well-characterized warm Neptune HAT-P-26 b. Despite its atmospheric metallicity being lower than that of Neptune \citep{Wakeford17,MacDonald19,Athano23}, its total metal content relative to its host star is similar to that of the Solar System's ice giants, being consistent with core accretion, and late pebble/planetesimal enrichment with weak core erosion. We constrain the mass of its core between 8.5 $M_{\oplus}$ (CMF = 0.45) and 14.4 $M_{\oplus}$ (CMF = 0.78). Our interior retrieval, which includes atmospheric metallicity, obtains better constraints on the core mass and the bulk metal mass fraction than previous estimates with only mass and radius data from \cite{Hartmann11}. 

Finally, we summarize our comparison between GASTLI and other interior models for exoplanets, and the effect of different assumptions on mass-radius relations and thermal evolution:

\begin{itemize}
    \item[-] Not including the contribution of the atmosphere from $\sim$1 bar to the transit pressure (20 mbar) underestimates the radius of Jupiter-mass planets by 2.5\%. However, for Neptune-mass planets, this corresponds to a 10-12\% of their radius, as shown by our comparison with models by \cite{Muller21}.
    \item[-] The \cite{Fortney07} (F07) models tend to overestimate the radius of Jupiter-mass planets by 5\% compared to our models and \cite{Muller21} (MH21). This may be caused by the use of the SCvH95 EOS \citep{Saumon95}, in contrast to the \cite{Chabrier21} (CD21) and \cite{Chabrier19} EOS. This is in agreement with a previous comparison by \cite{HG23}. These differences in density entail a change in CMF of 0.05-0.07, and 0.10 in envelope metal mass fraction. 
    \item[-] We observe that an enhancement of the atmospheric opacity can delay the thermal cooling of the interior, in agreement with previous work \citep{LF14,Poser19}. This produces a more luminous planet for the same age. We reproduce the effect of clouds in internal temperature evolution by enhancing the atmospheric metallicity in our clear models from solar to 10-100 $\times$ solar, which is compatible with previous works \citep{Linder18,Poser19,Poser24}.
    \item[-] For a constant mass, equilibrium temperature, and metallicity, a change in C/O ratio from 0.55 (solar) to 0.10 can increase the planet radius by 2\% for planets younger than 100 Myr. At these early stages of planet evolution, the choice of a hot or a cold start can also impact the planet radius.
\end{itemize}

Non-adiabatic temperature profiles caused by compositional gradients, as well as dilute cores, are required to explain the wealth of data from the Solar System's gas and ice giants. Upcoming observations with JWST \citep{Beichman14,Hellard20,Akinsanmi20}, PLATO \citep{plato_ref}, and Ariel \citep{ariel_ref} will provide data with unprecedented precision as well as new observables, such as Love numbers, for exoplanets. More detailed interior models of exoplanets are essential to interpret new data. GASTLI lays the groundwork for further development of open-source interior models, which will be the scope of future work.

\begin{acknowledgements}
      We thank Y. Miguel for sharing the re-formatted EOS tables for silicate (dry sand) from the SESAME database. We acknowledge the support of the Data Science Group at the Max Planck Institute for Astronomy (MPIA) and especially I. Momcheva, and M. Fouesneau for their invaluable assistance in optimizing the software, and developing the Python package for this research paper. L. A. thanks M. Deleuil, O. Mousis, A. Aguichine and B. Brugger for their contributions to the development of Marseille's Super-Earth Interior (MSEI) model, on which GASTLI's development is based.
\end{acknowledgements}

%
%


\bibliographystyle{aa}           
\bibliography{references}      

\begin{thebibliography}{158}
\expandafter\ifx\csname natexlab\endcsname\relax\def\natexlab#1{#1}\fi

\bibitem[{{A-thano} {et~al.}(2023){A-thano}, {Awiphan}, {Jiang}, {Kerins}, {Priyadarshi}, {McDonald}, {Joshi}, {Chulikorn}, {Hayes}, {Charles}, {Huang}, {Rattanamala}, {Yeh}, \& {Dhillon}}]{Athano23}
{A-thano}, N., {Awiphan}, S., {Jiang}, I.-G., {et~al.} 2023, \aj, 166, 223

\bibitem[{{Acu{\~n}a} {et~al.}(2023){Acu{\~n}a}, {Deleuil}, \& {Mousis}}]{Acuna23}
{Acu{\~n}a}, L., {Deleuil}, M., \& {Mousis}, O. 2023, \aap, 677, A14

\bibitem[{{Acu{\~n}a} {et~al.}(2021){Acu{\~n}a}, {Deleuil}, {Mousis}, {Marcq}, {Levesque}, \& {Aguichine}}]{Acuna21}
{Acu{\~n}a}, L., {Deleuil}, M., {Mousis}, O., {et~al.} 2021, \aap, 647, A53

\bibitem[{{Acu{\~n}a} {et~al.}(2022){Acu{\~n}a}, {Lopez}, {Morel}, {Deleuil}, {Mousis}, {Aguichine}, {Marcq}, \& {Santerne}}]{Acuna22}
{Acu{\~n}a}, L., {Lopez}, T.~A., {Morel}, T., {et~al.} 2022, \aap, 660, A102

\bibitem[{{Aguichine} {et~al.}(2021){Aguichine}, {Mousis}, {Deleuil}, \& {Marcq}}]{Aguichine21}
{Aguichine}, A., {Mousis}, O., {Deleuil}, M., \& {Marcq}, E. 2021, \apj, 914, 84

\bibitem[{{Ag{\'u}ndez} {et~al.}(2014){Ag{\'u}ndez}, {Venot}, {Selsis}, \& {Iro}}]{Agundez14}
{Ag{\'u}ndez}, M., {Venot}, O., {Selsis}, F., \& {Iro}, N. 2014, \apj, 781, 68

\bibitem[{{Akinsanmi} {et~al.}(2020){Akinsanmi}, {Barros}, {Santos}, {Oshagh}, \& {Serrano}}]{Akinsanmi20}
{Akinsanmi}, B., {Barros}, S.~C.~C., {Santos}, N.~C., {Oshagh}, M., \& {Serrano}, L.~M. 2020, \mnras, 497, 3484

\bibitem[{{Ali-Dib} \& {Lakhlani}(2018)}]{Alidib18}
{Ali-Dib}, M. \& {Lakhlani}, G. 2018, \mnras, 473, 1325

\bibitem[{{Arp} {et~al.}(1984){Arp}, {Persichetti}, \& {Chen}}]{Arp84}
{Arp}, V., {Persichetti}, J.~M., \& {Chen}, G.~B. 1984, ASME Journal of Fluids Engineering, 106, 193

\bibitem[{{Atreya} {et~al.}(2022){Atreya}, {Crida}, {Guillot}, {Li}, {Lunine}, {Madhusudhan}, {Mousis}, \& {Wong}}]{Atreya22}
{Atreya}, S.~K., {Crida}, A., {Guillot}, T., {et~al.} 2022, arXiv e-prints, arXiv:2205.06914

\bibitem[{{Baraffe} {et~al.}(2008){Baraffe}, {Chabrier}, \& {Barman}}]{Baraffe08}
{Baraffe}, I., {Chabrier}, G., \& {Barman}, T. 2008, \aap, 482, 315

\bibitem[{{Baraffe} {et~al.}(2003){Baraffe}, {Chabrier}, {Barman}, {Allard}, \& {Hauschildt}}]{Baraffe2003}
{Baraffe}, I., {Chabrier}, G., {Barman}, T.~S., {Allard}, F., \& {Hauschildt}, P.~H. 2003, \aap, 402, 701

\bibitem[{{Barros} {et~al.}(2022){Barros}, {Akinsanmi}, {Bou{\'e}}, {Smith}, {Laskar}, {Ulmer-Moll}, {Lillo-Box}, {Queloz}, {Cameron}, {Sousa}, {Ehrenreich}, {Hooton}, {Bruno}, {Demory}, {Correia}, {Demangeon}, {Wilson}, {Bonfanti}, {Hoyer}, {Alibert}, {Alonso}, {Escud{\'e}}, {Barbato}, {B{\'a}rczy}, {Barrado}, {Baumjohann}, {Beck}, {Beck}, {Benz}, {Bergomi}, {Billot}, {Bonfils}, {Bouchy}, {Brandeker}, {Broeg}, {Cabrera}, {Cessa}, {Charnoz}, {Damme}, {Davies}, {Deleuil}, {Deline}, {Delrez}, {Erikson}, {Fortier}, {Fossati}, {Fridlund}, {Gandolfi}, {Mu{\~n}oz}, {Gillon}, {G{\"u}del}, {Isaak}, {Heng}, {Kiss}, {des Etangs}, {Lendl}, {Lovis}, {Magrin}, {Nascimbeni}, {Maxted}, {Olofsson}, {Ottensamer}, {Pagano}, {Pall{\'e}}, {Parviainen}, {Peter}, {Piotto}, {Pollacco}, {Ragazzoni}, {Rando}, {Rauer}, {Ribas}, {Santos}, {Scandariato}, {S{\'e}gransan}, {Simon}, {Steller}, {Szab{\'o}}, {Thomas}, {Udry}, {Ulmer}, {Van Grootel}, \& {Walton}}]{Barros22}
{Barros}, S.~C.~C., {Akinsanmi}, B., {Bou{\'e}}, G., {et~al.} 2022, \aap, 657, A52

\bibitem[{{Batygin} \& {Stevenson}(2010)}]{Batygin10}
{Batygin}, K. \& {Stevenson}, D.~J. 2010, \apjl, 714, L238

\bibitem[{{Baudino} {et~al.}(2017){Baudino}, {Molli{\`e}re}, {Venot}, {Tremblin}, {B{\'e}zard}, \& {Lagage}}]{Baudino17}
{Baudino}, J.-L., {Molli{\`e}re}, P., {Venot}, O., {et~al.} 2017, \apj, 850, 150

\bibitem[{{Baumeister} \& {Tosi}(2023)}]{Baumeister23}
{Baumeister}, P. \& {Tosi}, N. 2023, \aap, 676, A106

\bibitem[{{Beichman} {et~al.}(2014){Beichman}, {Benneke}, {Knutson}, {Smith}, {Lagage}, {Dressing}, {Latham}, {Lunine}, {Birkmann}, {Ferruit}, {Giardino}, {Kempton}, {Carey}, {Krick}, {Deroo}, {Mandell}, {Ressler}, {Shporer}, {Swain}, {Vasisht}, {Ricker}, {Bouwman}, {Crossfield}, {Greene}, {Howell}, {Christiansen}, {Ciardi}, {Clampin}, {Greenhouse}, {Sozzetti}, {Goudfrooij}, {Hines}, {Keyes}, {Lee}, {McCullough}, {Robberto}, {Stansberry}, {Valenti}, {Rieke}, {Rieke}, {Fortney}, {Bean}, {Kreidberg}, {Ehrenreich}, {Deming}, {Albert}, {Doyon}, \& {Sing}}]{Beichman14}
{Beichman}, C., {Benneke}, B., {Knutson}, H., {et~al.} 2014, \pasp, 126, 1134

\bibitem[{{Benneke} {et~al.}(2024){Benneke}, {Roy}, {Coulombe}, {Radica}, {Piaulet}, {Ahrer}, {Pierrehumbert}, {Krissansen-Totton}, {Schlichting}, {Hu}, {Yang}, {Christie}, {Thorngren}, {Young}, {Pelletier}, {Knutson}, {Miguel}, {Evans-Soma}, {Dorn}, {Gagnebin}, {Fortney}, {Komacek}, {MacDonald}, {Raul}, {Cloutier}, {Acuna}, {Lafreni{\`e}re}, {Cadieux}, {Doyon}, {Welbanks}, \& {Allart}}]{Benneke24}
{Benneke}, B., {Roy}, P.-A., {Coulombe}, L.-P., {et~al.} 2024, arXiv e-prints, arXiv:2403.03325

\bibitem[{{Berardo} \& {de Wit}(2022)}]{Berardo22}
{Berardo}, D. \& {de Wit}, J. 2022, \apj, 935, 178

\bibitem[{{Bernab{\`o}} {et~al.}(2024){Bernab{\`o}}, {Csizmadia}, {Smith}, {Rauer}, {Hatzes}, {Esposito}, {Gandolfi}, \& {Cabrera}}]{Bernabo24}
{Bernab{\`o}}, L.~M., {Csizmadia}, S., {Smith}, A.~M.~S., {et~al.} 2024, \aap, 684, A78

\bibitem[{{Bloot} {et~al.}(2023){Bloot}, {Miguel}, {Bazot}, \& {Howard}}]{Bloot23}
{Bloot}, S., {Miguel}, Y., {Bazot}, M., \& {Howard}, S. 2023, \mnras, 523, 6282

\bibitem[{{Bodenheimer} {et~al.}(2001){Bodenheimer}, {Lin}, \& {Mardling}}]{Bodenheimer01}
{Bodenheimer}, P., {Lin}, D.~N.~C., \& {Mardling}, R.~A. 2001, \apj, 548, 466

\bibitem[{Brugger(2018)}]{Brugger_phd_thesis}
Brugger, B. 2018, PhD thesis, thèse de doctorat dirigée par Mousis, Olivier et Deleuil, Magali Astrophysique et Cosmologie Aix-Marseille 2018

\bibitem[{{Brugger} {et~al.}(2017){Brugger}, {Mousis}, {Deleuil}, \& {Deschamps}}]{Brugger17}
{Brugger}, B., {Mousis}, O., {Deleuil}, M., \& {Deschamps}, F. 2017, \apj, 850, 93

\bibitem[{{Brugger} {et~al.}(2016){Brugger}, {Mousis}, {Deleuil}, \& {Lunine}}]{Brugger16}
{Brugger}, B., {Mousis}, O., {Deleuil}, M., \& {Lunine}, J.~I. 2016, \apjl, 831, L16

\bibitem[{{Buhler} {et~al.}(2016){Buhler}, {Knutson}, {Batygin}, {Fulton}, {Fortney}, {Burrows}, \& {Wong}}]{Buhler16}
{Buhler}, P.~B., {Knutson}, H.~A., {Batygin}, K., {et~al.} 2016, \apj, 821, 26

\bibitem[{{Burrows} {et~al.}(2001){Burrows}, {Hubbard}, {Lunine}, \& {Liebert}}]{Burrows2001}
{Burrows}, A., {Hubbard}, W.~B., {Lunine}, J.~I., \& {Liebert}, J. 2001, Reviews of Modern Physics, 73, 719

\bibitem[{{Burrows} {et~al.}(1993){Burrows}, {Hubbard}, {Saumon}, \& {Lunine}}]{Burrows93}
{Burrows}, A., {Hubbard}, W.~B., {Saumon}, D., \& {Lunine}, J.~I. 1993, \apj, 406, 158

\bibitem[{{Burrows} {et~al.}(1997){Burrows}, {Marley}, {Hubbard}, {Lunine}, {Guillot}, {Saumon}, {Freedman}, {Sudarsky}, \& {Sharp}}]{Burrows97}
{Burrows}, A., {Marley}, M., {Hubbard}, W.~B., {et~al.} 1997, \apj, 491, 856

\bibitem[{{Burrows} {et~al.}(1995){Burrows}, {Saumon}, {Guillot}, {Hubbard}, \& {Lunine}}]{Burrows95}
{Burrows}, A., {Saumon}, D., {Guillot}, T., {Hubbard}, W.~B., \& {Lunine}, J.~I. 1995, \nat, 375, 299

\bibitem[{{Cavali{\'e}} {et~al.}(2023){Cavali{\'e}}, {Lunine}, \& {Mousis}}]{Cavalie23}
{Cavali{\'e}}, T., {Lunine}, J., \& {Mousis}, O. 2023, Nature Astronomy, 7, 678

\bibitem[{{Chabrier} \& {Debras}(2021)}]{Chabrier21}
{Chabrier}, G. \& {Debras}, F. 2021, \apj, 917, 4

\bibitem[{{Chabrier} {et~al.}(2019){Chabrier}, {Mazevet}, \& {Soubiran}}]{Chabrier19}
{Chabrier}, G., {Mazevet}, S., \& {Soubiran}, F. 2019, \apj, 872, 51

\bibitem[{{Chabrier} {et~al.}(1992){Chabrier}, {Saumon}, {Hubbard}, \& {Lunine}}]{Chabrier92}
{Chabrier}, G., {Saumon}, D., {Hubbard}, W.~B., \& {Lunine}, J.~I. 1992, \apj, 391, 817

\bibitem[{{Chen} \& {Rogers}(2016)}]{chen_rogers16}
{Chen}, H. \& {Rogers}, L.~A. 2016, \apj, 831, 180

\bibitem[{{Chen} {et~al.}(2023){Chen}, {Burrows}, {Sur}, \& {Arevalo}}]{Chen23}
{Chen}, Y.-X., {Burrows}, A., {Sur}, A., \& {Arevalo}, R.~T. 2023, \apj, 957, 36

\bibitem[{{Csizmadia} {et~al.}(2019){Csizmadia}, {Hellard}, \& {Smith}}]{Csizmadia19}
{Csizmadia}, S., {Hellard}, H., \& {Smith}, A.~M.~S. 2019, \aap, 623, A45

\bibitem[{{Deitrick} {et~al.}(2020){Deitrick}, {Mendon{\c{c}}a}, {Schroffenegger}, {Grimm}, {Tsai}, \& {Heng}}]{thor_20}
{Deitrick}, R., {Mendon{\c{c}}a}, J.~M., {Schroffenegger}, U., {et~al.} 2020, \apjs, 248, 30

\bibitem[{{Dorn} {et~al.}(2015){Dorn}, {Khan}, {Heng}, {Connolly}, {Alibert}, {Benz}, \& {Tackley}}]{Dorn15}
{Dorn}, C., {Khan}, A., {Heng}, K., {et~al.} 2015, \aap, 577, A83

\bibitem[{Duvall \& Taylor(1971)}]{Duval71}
Duvall \& Taylor. 1971, J. Composite Materials, 5, 130

\bibitem[{{Foreman-Mackey} {et~al.}(2013){Foreman-Mackey}, {Hogg}, {Lang}, \& {Goodman}}]{emcee}
{Foreman-Mackey}, D., {Hogg}, D.~W., {Lang}, D., \& {Goodman}, J. 2013, \pasp, 125, 306

\bibitem[{{Fortney} \& {Hubbard}(2003)}]{Fortney_Hubbard2003}
{Fortney}, J.~J. \& {Hubbard}, W.~B. 2003, \icarus, 164, 228

\bibitem[{{Fortney} \& {Hubbard}(2004)}]{Fortney04}
{Fortney}, J.~J. \& {Hubbard}, W.~B. 2004, \apj, 608, 1039

\bibitem[{{Fortney} {et~al.}(2007){Fortney}, {Marley}, \& {Barnes}}]{Fortney07}
{Fortney}, J.~J., {Marley}, M.~S., \& {Barnes}, J.~W. 2007, \apj, 659, 1661

\bibitem[{{Fortney} {et~al.}(2013){Fortney}, {Mordasini}, {Nettelmann}, {Kempton}, {Greene}, \& {Zahnle}}]{Fortney13}
{Fortney}, J.~J., {Mordasini}, C., {Nettelmann}, N., {et~al.} 2013, \apj, 775, 80

\bibitem[{{Fortney} {et~al.}(2020){Fortney}, {Visscher}, {Marley}, {Hood}, {Line}, {Thorngren}, {Freedman}, \& {Lupu}}]{Fortney20}
{Fortney}, J.~J., {Visscher}, C., {Marley}, M.~S., {et~al.} 2020, \aj, 160, 288

\bibitem[{{French} {et~al.}(2009){French}, {Mattsson}, {Nettelmann}, \& {Redmer}}]{French09}
{French}, M., {Mattsson}, T.~R., {Nettelmann}, N., \& {Redmer}, R. 2009, \prb, 79, 054107

\bibitem[{Gordon(1994)}]{Gordon_cea}
Gordon, S. 1994, Tech. rep. NASA Lewis Research Center.

\bibitem[{{Grimm} {et~al.}(2018){Grimm}, {Demory}, {Gillon}, {Dorn}, {Agol}, {Burdanov}, {Delrez}, {Sestovic}, {Triaud}, {Turbet}, {Bolmont}, {Caldas}, {de Wit}, {Jehin}, {Leconte}, {Raymond}, {Van Grootel}, {Burgasser}, {Carey}, {Fabrycky}, {Heng}, {Hernandez}, {Ingalls}, {Lederer}, {Selsis}, \& {Queloz}}]{Grimm18}
{Grimm}, S.~L., {Demory}, B.-O., {Gillon}, M., {et~al.} 2018, \aap, 613, A68

\bibitem[{{Guillot}(2010)}]{Guillot10}
{Guillot}, T. 2010, \aap, 520, A27

\bibitem[{{Gupta} {et~al.}(2022){Gupta}, {Atreya}, {Steffes}, {Fletcher}, {Guillot}, {Allison}, {Bolton}, {Helled}, {Levin}, {Li}, {Lunine}, {Miguel}, {Orton}, {Hunter Waite}, \& {Withers}}]{Gupta22}
{Gupta}, P., {Atreya}, S.~K., {Steffes}, P.~G., {et~al.} 2022, \psj, 3, 159

\bibitem[{{Haldemann} {et~al.}(2020){Haldemann}, {Alibert}, {Mordasini}, \& {Benz}}]{Haldemann20}
{Haldemann}, J., {Alibert}, Y., {Mordasini}, C., \& {Benz}, W. 2020, \aap, 643, A105

\bibitem[{{Haldemann} {et~al.}(2024){Haldemann}, {Dorn}, {Venturini}, {Alibert}, \& {Benz}}]{haldemann24}
{Haldemann}, J., {Dorn}, C., {Venturini}, J., {Alibert}, Y., \& {Benz}, W. 2024, \aap, 681, A96

\bibitem[{{Hardy} {et~al.}(2017){Hardy}, {Harrington}, {Hardin}, {Madhusudhan}, {Loredo}, {Challener}, {Foster}, {Cubillos}, \& {Blecic}}]{Hardy17}
{Hardy}, R.~A., {Harrington}, J., {Hardin}, M.~R., {et~al.} 2017, \apj, 836, 143

\bibitem[{{Hartman} {et~al.}(2011){Hartman}, {Bakos}, {Kipping}, {Torres}, {Kov{\'a}cs}, {Noyes}, {Latham}, {Howard}, {Fischer}, {Johnson}, {Marcy}, {Isaacson}, {Quinn}, {Buchhave}, {B{\'e}ky}, {Sasselov}, {Stefanik}, {Esquerdo}, {Everett}, {Perumpilly}, {L{\'a}z{\'a}r}, {Papp}, \& {S{\'a}ri}}]{Hartmann11}
{Hartman}, J.~D., {Bakos}, G.~{\'A}., {Kipping}, D.~M., {et~al.} 2011, \apj, 728, 138

\bibitem[{{Hellard} {et~al.}(2020){Hellard}, {Csizmadia}, {Padovan}, {Sohl}, \& {Rauer}}]{Hellard20}
{Hellard}, H., {Csizmadia}, S., {Padovan}, S., {Sohl}, F., \& {Rauer}, H. 2020, \apj, 889, 66

\bibitem[{{Helled}(2023)}]{Helled23}
{Helled}, R. 2023, \aap, 675, L8

\bibitem[{{Helled} {et~al.}(2011){Helled}, {Anderson}, {Podolak}, \& {Schubert}}]{Helled11}
{Helled}, R., {Anderson}, J.~D., {Podolak}, M., \& {Schubert}, G. 2011, \apj, 726, 15

\bibitem[{{Helled} \& {Morbidelli}(2021)}]{HM21}
{Helled}, R. \& {Morbidelli}, A. 2021, in ExoFrontiers; Big Questions in Exoplanetary Science, ed. N.~{Madhusudhan}, 12--1

\bibitem[{{Helled} \& {Stevenson}(2017)}]{Helled17}
{Helled}, R. \& {Stevenson}, D. 2017, \apjl, 840, L4

\bibitem[{{Helled} \& {Stevenson}(2024)}]{Helled24_rev}
{Helled}, R. \& {Stevenson}, D.~J. 2024, arXiv e-prints, arXiv:2403.11657

\bibitem[{{Howard} \& {Guillot}(2023)}]{HG23}
{Howard}, S. \& {Guillot}, T. 2023, arXiv e-prints, arXiv:2302.07902

\bibitem[{{Howard} {et~al.}(2023){Howard}, {Guillot}, {Bazot}, {Miguel}, {Stevenson}, {Galanti}, {Kaspi}, {Hubbard}, {Militzer}, {Helled}, {Nettelmann}, {Idini}, \& {Bolton}}]{Howard23}
{Howard}, S., {Guillot}, T., {Bazot}, M., {et~al.} 2023, \aap, 672, A33

\bibitem[{{Huang} {et~al.}(2022){Huang}, {Rice}, \& {Steffen}}]{magrathea}
{Huang}, C., {Rice}, D.~R., \& {Steffen}, J.~H. 2022, \mnras, 513, 5256

\bibitem[{{Hubbard}(1977)}]{Hubbard77}
{Hubbard}, W.~B. 1977, \icarus, 30, 305

\bibitem[{{Hubeny} \& {Burrows}(2007)}]{Hubeny07}
{Hubeny}, I. \& {Burrows}, A. 2007, \apj, 669, 1248

\bibitem[{{Ibgui} {et~al.}(2010){Ibgui}, {Burrows}, \& {Spiegel}}]{Ibgui10}
{Ibgui}, L., {Burrows}, A., \& {Spiegel}, D.~S. 2010, \apj, 713, 751

\bibitem[{{Ibgui} {et~al.}(2011){Ibgui}, {Spiegel}, \& {Burrows}}]{Ibgui11}
{Ibgui}, L., {Spiegel}, D.~S., \& {Burrows}, A. 2011, \apj, 727, 75

\bibitem[{{Irwin} {et~al.}(2021){Irwin}, {Dobinson}, {James}, {Toledo}, {Teanby}, {Fletcher}, {Orton}, \& {P{\'e}rez-Hoyos}}]{Irwin21}
{Irwin}, P. G.~J., {Dobinson}, J., {James}, A., {et~al.} 2021, \icarus, 357, 114277

\bibitem[{{Irwin} {et~al.}(2019){Irwin}, {Toledo}, {Braude}, {Bacon}, {Weilbacher}, {Teanby}, {Fletcher}, \& {Orton}}]{Irwin19}
{Irwin}, P. G.~J., {Toledo}, D., {Braude}, A.~S., {et~al.} 2019, \icarus, 331, 69

\bibitem[{{Jin} {et~al.}(2014){Jin}, {Mordasini}, {Parmentier}, {van Boekel}, {Henning}, \& {Ji}}]{Jin14}
{Jin}, S., {Mordasini}, C., {Parmentier}, V., {et~al.} 2014, \apj, 795, 65

\bibitem[{{Karkoschka} \& {Tomasko}(2011)}]{Karkoschka11}
{Karkoschka}, E. \& {Tomasko}, M.~G. 2011, \icarus, 211, 780

\bibitem[{{Komacek} {et~al.}(2022){Komacek}, {Gao}, {Thorngren}, {May}, \& {Tan}}]{Komacek22}
{Komacek}, T.~D., {Gao}, P., {Thorngren}, D.~P., {May}, E.~M., \& {Tan}, X. 2022, \apjl, 941, L40

\bibitem[{{Kramm} {et~al.}(2011){Kramm}, {Nettelmann}, {Redmer}, \& {Stevenson}}]{Kramm11}
{Kramm}, U., {Nettelmann}, N., {Redmer}, R., \& {Stevenson}, D.~J. 2011, \aap, 528, A18

\bibitem[{{Kurucz}(1993)}]{Kurucz93}
{Kurucz}, R.~L. 1993, {SYNTHE spectrum synthesis programs and line data}

\bibitem[{{Leconte} \& {Chabrier}(2012)}]{Leconte12}
{Leconte}, J. \& {Chabrier}, G. 2012, \aap, 540, A20

\bibitem[{{Leconte} {et~al.}(2010){Leconte}, {Chabrier}, {Baraffe}, \& {Levrard}}]{Leconte10}
{Leconte}, J., {Chabrier}, G., {Baraffe}, I., \& {Levrard}, B. 2010, \aap, 516, A64

\bibitem[{Li {et~al.}(2024)Li, Allison, Atreya, Brueshaber, Fletcher, Guillot, Li, Lunine, Miguel, Orton, Steffes, Waite, Wong, Levin, \& Bolton}]{Li24}
Li, C., Allison, M., Atreya, S., {et~al.} 2024, Icarus, 116028

\bibitem[{{Li} {et~al.}(2020){Li}, {Ingersoll}, {Bolton}, {Levin}, {Janssen}, {Atreya}, {Lunine}, {Steffes}, {Brown}, {Guillot}, {Allison}, {Arballo}, {Bellotti}, {Adumitroaie}, {Gulkis}, {Hodges}, {Li}, {Misra}, {Orton}, {Oyafuso}, {Santos-Costa}, {Waite}, \& {Zhang}}]{Li20}
{Li}, C., {Ingersoll}, A., {Bolton}, S., {et~al.} 2020, Nature Astronomy, 4, 609

\bibitem[{{Li} {et~al.}(2018){Li}, {Oyafuso}, {Brown}, {Adumitroaie}, {Bellotti}, {Arballo}, {Aglyamov}, {Allison}, {Atreya}, {Gulkis}, {Ingersoll}, {Janssen}, {Lunine}, {Misra}, {Orton}, {Santos-Costa}, {Sarkissian}, {Steffes}, {Levin}, \& {Bolton}}]{Li18}
{Li}, C., {Oyafuso}, F.~A., {Brown}, S.~T., {et~al.} 2018, in AGU Fall Meeting Abstracts, Vol. 2018, P33F--3884

\bibitem[{{Linder} {et~al.}(2019){Linder}, {Mordasini}, {Molli{\`e}re}, {Marleau}, {Malik}, {Quanz}, \& {Meyer}}]{Linder18}
{Linder}, E.~F., {Mordasini}, C., {Molli{\`e}re}, P., {et~al.} 2019, \aap, 623, A85

\bibitem[{{Lodders}(2003)}]{Lodders03}
{Lodders}, K. 2003, \apj, 591, 1220

\bibitem[{{Lopez} \& {Fortney}(2014)}]{LF14}
{Lopez}, E.~D. \& {Fortney}, J.~J. 2014, \apj, 792, 1

\bibitem[{{Love}(1911)}]{Love11}
{Love}, A.~E.~H. 1911, {Some Problems of Geodynamics}

\bibitem[{{Lozovsky} {et~al.}(2017){Lozovsky}, {Helled}, {Rosenberg}, \& {Bodenheimer}}]{Lozovsky17}
{Lozovsky}, M., {Helled}, R., {Rosenberg}, E.~D., \& {Bodenheimer}, P. 2017, \apj, 836, 227

\bibitem[{Lyon(1992)}]{sesame}
Lyon, S.~P. 1992, Los Alamos National Laboratory report LA-UR-92-3407

\bibitem[{{MacDonald} \& {Madhusudhan}(2019)}]{MacDonald19}
{MacDonald}, R.~J. \& {Madhusudhan}, N. 2019, \mnras, 486, 1292

\bibitem[{{MacKenzie} {et~al.}(2023){MacKenzie}, {Grenfell}, {Baumeister}, {Tosi}, {Cabrera}, \& {Rauer}}]{MacKenzie23}
{MacKenzie}, J., {Grenfell}, J.~L., {Baumeister}, P., {et~al.} 2023, \aap, 671, A65

\bibitem[{{Madhusudhan} {et~al.}(2020){Madhusudhan}, {Nixon}, {Welbanks}, {Piette}, \& {Booth}}]{Madhu20}
{Madhusudhan}, N., {Nixon}, M.~C., {Welbanks}, L., {Piette}, A. A.~A., \& {Booth}, R.~A. 2020, \apjl, 891, L7

\bibitem[{{Mahaffy} {et~al.}(2000){Mahaffy}, {Niemann}, {Alpert}, {Atreya}, {Demick}, {Donahue}, {Harpold}, \& {Owen}}]{Mahaffy00}
{Mahaffy}, P.~R., {Niemann}, H.~B., {Alpert}, A., {et~al.} 2000, \jgr, 105, 15061

\bibitem[{{Mankovich} \& {Fortney}(2020)}]{Mankovich20}
{Mankovich}, C.~R. \& {Fortney}, J.~J. 2020, \apj, 889, 51

\bibitem[{{Mankovich} \& {Fuller}(2021)}]{Mankovich21}
{Mankovich}, C.~R. \& {Fuller}, J. 2021, Nature Astronomy, 5, 1103

\bibitem[{{Marley} {et~al.}(2007){Marley}, {Fortney}, {Hubickyj}, {Bodenheimer}, \& {Lissauer}}]{Marley07}
{Marley}, M.~S., {Fortney}, J.~J., {Hubickyj}, O., {Bodenheimer}, P., \& {Lissauer}, J.~J. 2007, \apj, 655, 541

\bibitem[{{Marley} \& {Robinson}(2015)}]{Marley15}
{Marley}, M.~S. \& {Robinson}, T.~D. 2015, \araa, 53, 279

\bibitem[{{Marley} {et~al.}(2012){Marley}, {Saumon}, {Cushing}, {Ackerman}, {Fortney}, \& {Freedman}}]{Marley12}
{Marley}, M.~S., {Saumon}, D., {Cushing}, M., {et~al.} 2012, \apj, 754, 135

\bibitem[{{Marley} {et~al.}(2021){Marley}, {Saumon}, {Visscher}, {Lupu}, {Freedman}, {Morley}, {Fortney}, {Seay}, {Smith}, {Teal}, \& {Wang}}]{Marley21}
{Marley}, M.~S., {Saumon}, D., {Visscher}, C., {et~al.} 2021, \apj, 920, 85

\bibitem[{{Mazevet} {et~al.}(2019){Mazevet}, {Licari}, {Chabrier}, \& {Potekhin}}]{Mazevet19}
{Mazevet}, S., {Licari}, A., {Chabrier}, G., \& {Potekhin}, A.~Y. 2019, \aap, 621, A128

\bibitem[{McBride(1996)}]{Mcbride_cea}
McBride, B. J.~G. 1996, Tech. rep. NASA Lewis Research Center.

\bibitem[{{Mercer} \& {Stamatellos}(2020)}]{Mercer20}
{Mercer}, A. \& {Stamatellos}, D. 2020, \aap, 633, A116

\bibitem[{{Miguel} {et~al.}(2022){Miguel}, {Bazot}, {Guillot}, {Howard}, {Galanti}, {Kaspi}, {Hubbard}, {Militzer}, {Helled}, {Atreya}, {Connerney}, {Durante}, {Kulowski}, {Lunine}, {Stevenson}, \& {Bolton}}]{Miguel22}
{Miguel}, Y., {Bazot}, M., {Guillot}, T., {et~al.} 2022, \aap, 662, A18

\bibitem[{{Miguel} \& {Vazan}(2023)}]{MV23}
{Miguel}, Y. \& {Vazan}, A. 2023, Remote Sensing, 15, 681

\bibitem[{{Militzer} {et~al.}(2019){Militzer}, {Wahl}, \& {Hubbard}}]{Militzer19}
{Militzer}, B., {Wahl}, S., \& {Hubbard}, W.~B. 2019, \apj, 879, 78

\bibitem[{{Molli{\`e}re} {et~al.}(2017){Molli{\`e}re}, {van Boekel}, {Bouwman}, {Henning}, {Lagage}, \& {Min}}]{Molliere17}
{Molli{\`e}re}, P., {van Boekel}, R., {Bouwman}, J., {et~al.} 2017, \aap, 600, A10

\bibitem[{{Molli{\`e}re} {et~al.}(2015){Molli{\`e}re}, {van Boekel}, {Dullemond}, {Henning}, \& {Mordasini}}]{Molliere15}
{Molli{\`e}re}, P., {van Boekel}, R., {Dullemond}, C., {Henning}, T., \& {Mordasini}, C. 2015, \apj, 813, 47

\bibitem[{{Morley} {et~al.}(2024){Morley}, {Mukherjee}, {Marley}, {Fortney}, {Visscher}, {Lupu}, {Gharib-Nezhad}, {Thorngren}, {Freedman}, \& {Batalha 7}}]{Morley24}
{Morley}, C.~V., {Mukherjee}, S., {Marley}, M.~S., {et~al.} 2024, arXiv e-prints, arXiv:2402.00758

\bibitem[{{Mousis} {et~al.}(2020){Mousis}, {Deleuil}, {Aguichine}, {Marcq}, {Naar}, {Aguirre}, {Brugger}, \& {Gon{\c{c}}alves}}]{Mousis20}
{Mousis}, O., {Deleuil}, M., {Aguichine}, A., {et~al.} 2020, \apjl, 896, L22

\bibitem[{{M{\"u}ller} \& {Helled}(2021)}]{Muller21}
{M{\"u}ller}, S. \& {Helled}, R. 2021, \mnras, 507, 2094

\bibitem[{{M{\"u}ller} \& {Helled}(2023)}]{Muller22}
{M{\"u}ller}, S. \& {Helled}, R. 2023, \aap, 669, A24

\bibitem[{Müller \& Helled(2024)}]{MH24}
Müller, S. \& Helled, R. 2024, Can Jupiter's atmospheric metallicity be different from the deep interior?

\bibitem[{{Nettelmann} {et~al.}(2008){Nettelmann}, {Holst}, {Kietzmann}, {French}, {Redmer}, \& {Blaschke}}]{Nettelmann08}
{Nettelmann}, N., {Holst}, B., {Kietzmann}, A., {et~al.} 2008, \apj, 683, 1217

\bibitem[{{Neuenschwander} {et~al.}(2021){Neuenschwander}, {Helled}, {Movshovitz}, \& {Fortney}}]{Neuenschwander21}
{Neuenschwander}, B.~A., {Helled}, R., {Movshovitz}, N., \& {Fortney}, J.~J. 2021, \apj, 910, 38

\bibitem[{{Ohno} \& {Fortney}(2023)}]{Ohno23}
{Ohno}, K. \& {Fortney}, J.~J. 2023, \apj, 946, 18

\bibitem[{{Orton}(1975)}]{Orton75}
{Orton}, G.~S. 1975, \icarus, 26, 125

\bibitem[{{Otegi} {et~al.}(2020){Otegi}, {Dorn}, {Helled}, {Bouchy}, {Haldemann}, \& {Alibert}}]{Otegi20}
{Otegi}, J.~F., {Dorn}, C., {Helled}, R., {et~al.} 2020, \aap, 640, A135

\bibitem[{{Owen} \& {Wu}(2013)}]{OW13}
{Owen}, J.~E. \& {Wu}, Y. 2013, \apj, 775, 105

\bibitem[{{Panwar} {et~al.}(2022){Panwar}, {D{\'e}sert}, {Todorov}, {Bean}, {Stevenson}, {Huitson}, {Fortney}, \& {Bergmann}}]{Panwar22}
{Panwar}, V., {D{\'e}sert}, J.-M., {Todorov}, K.~O., {et~al.} 2022, \mnras, 510, 3236

\bibitem[{{Paxton} {et~al.}(2011){Paxton}, {Bildsten}, {Dotter}, {Herwig}, {Lesaffre}, \& {Timmes}}]{Paxton11}
{Paxton}, B., {Bildsten}, L., {Dotter}, A., {et~al.} 2011, \apjs, 192, 3

\bibitem[{{Paxton} {et~al.}(2013){Paxton}, {Cantiello}, {Arras}, {Bildsten}, {Brown}, {Dotter}, {Mankovich}, {Montgomery}, {Stello}, {Timmes}, \& {Townsend}}]{Paxton13}
{Paxton}, B., {Cantiello}, M., {Arras}, P., {et~al.} 2013, \apjs, 208, 4

\bibitem[{{Paxton} {et~al.}(2015){Paxton}, {Marchant}, {Schwab}, {Bauer}, {Bildsten}, {Cantiello}, {Dessart}, {Farmer}, {Hu}, {Langer}, {Townsend}, {Townsley}, \& {Timmes}}]{Paxton15}
{Paxton}, B., {Marchant}, P., {Schwab}, J., {et~al.} 2015, \apjs, 220, 15

\bibitem[{{Paxton} {et~al.}(2018){Paxton}, {Schwab}, {Bauer}, {Bildsten}, {Blinnikov}, {Duffell}, {Farmer}, {Goldberg}, {Marchant}, {Sorokina}, {Thoul}, {Townsend}, \& {Timmes}}]{Paxton18}
{Paxton}, B., {Schwab}, J., {Bauer}, E.~B., {et~al.} 2018, \apjs, 234, 34

\bibitem[{{Paxton} {et~al.}(2019){Paxton}, {Smolec}, {Schwab}, {Gautschy}, {Bildsten}, {Cantiello}, {Dotter}, {Farmer}, {Goldberg}, {Jermyn}, {Kanbur}, {Marchant}, {Thoul}, {Townsend}, {Wolf}, {Zhang}, \& {Timmes}}]{Paxton19}
{Paxton}, B., {Smolec}, R., {Schwab}, J., {et~al.} 2019, \apjs, 243, 10

\bibitem[{{Pearl} \& {Conrath}(1991)}]{Pearl91}
{Pearl}, J.~C. \& {Conrath}, B.~J. 1991, \jgr, 96, 18921

\bibitem[{{Peebles}(1964)}]{Pebbles64}
{Peebles}, P.~J.~E. 1964, \apj, 140, 328

\bibitem[{{Perna} {et~al.}(2010){Perna}, {Menou}, \& {Rauscher}}]{Perna10}
{Perna}, R., {Menou}, K., \& {Rauscher}, E. 2010, \apj, 724, 313

\bibitem[{{Podolak} {et~al.}(2019){Podolak}, {Helled}, \& {Schubert}}]{Podolak19}
{Podolak}, M., {Helled}, R., \& {Schubert}, G. 2019, \mnras, 487, 2653

\bibitem[{{Poser} {et~al.}(2019){Poser}, {Nettelmann}, \& {Redmer}}]{Poser19}
{Poser}, A.~J., {Nettelmann}, N., \& {Redmer}, R. 2019, Atmosphere, 10, 664

\bibitem[{{Poser} \& {Redmer}(2024)}]{Poser24}
{Poser}, A.~J. \& {Redmer}, R. 2024, arXiv e-prints, arXiv:2402.19466

\bibitem[{{Rauer} {et~al.}(2024){Rauer}, {Aerts}, {Cabrera}, {Deleuil}, {Erikson}, {Gizon}, {Goupil}, {Heras}, {Lorenzo-Alvarez}, {Marliani}, {Martin-Garcia}, {Mas-Hesse}, {O'Rourke}, {Osborn}, {Pagano}, {Piotto}, {Pollacco}, {Ragazzoni}, {Ramsay}, {Udry}, {Appourchaux}, {Benz}, {Brandeker}, {G{\"u}del}, {Janot-Pacheco}, {Kabath}, {Kjeldsen}, {Min}, {Santos}, {Smith}, {Suarez}, {Werner}, {Aboudan}, {Abreu}, {Acu{\~n}a}, {Adams}, {Adibekyan}, {Affer}, {Agneray}, {Agnor}, {Aguirre B{\o}rsen-Koch}, {Ahmed}, {Aigrain}, {Al-Bahlawan}, {Alcacera Gil}, {Alei}, {Alencar}, {Alexander}, {Alfonso-Garz{\'o}n}, {Alibert}, {Allende Prieto}, {Almeida}, {Alonso Sobrino}, {Altavilla}, {Althaus}, {Alonso Alvarez Trujillo}, {Amarsi}, {Ammler-von Eiff}, {Am{\^o}res}, {Andrade}, {Antoniadis-Karnavas}, {Ant{\'o}nio}, {Aparicio del Moral}, {Appolloni}, {Arena}, {Armstrong}, {Aroca Aliaga}, {Asplund}, {Audenaert}, {Auricchio}, {Avelino}, {Baeke}, {Bailli{\'e}}, {Balado}, {Balestra}, {Ball}, {Ballans}, {Ballot}, {Barban}, {Barbary},
  {Barbieri}, {Barcel{\'o} Forteza}, {Barker}, {Barklem}, {Barnes}, {Barrado Navascues}, {Barragan}, {Baruteau}, {Basu}, {Baudin}, {Baumeister}, {Bayliss}, {Bazot}, {Beck}, {Bedding}, {Belkacem}, {Bellinger}, {Benatti}, {Benomar}, {B{\'e}rard}, {Bergemann}, {Bergomi}, {Bernardo}, {Biazzo}, {Bignamini}, {Bigot}, {Billot}, {Binet}, {Biondi}, {Biondi}, {Birch}, {Bitsch}, {Bluhm Ceballos}, {B{\'o}di}, {Bogn{\'a}r}, {Boisse}, {Bolmont}, {Bonanno}, {Bonavita}, {Bonfanti}, {Bonfils}, {Bonito}, {Bonomo}, {B{\"o}rner}, {Boro Saikia}, {Borreguero Mart{\'\i}n}, {Borsa}, {Borsato}, {Bossini}, {Bouchy}, {Bou{\'e}}, {Boufleur}, {Boumier}, {Bourrier}, {Bowman}, {Bozzo}, {Bradley}, {Bray}, {Bressan}, {Breton}, {Brienza}, {Brito}, {Brogi}, {Brown}, {Brown}, {Brun}, {Bruno}, {Bruns}, {Buchhave}, {Bugnet}, {Buldgen}, {Burgess}, {Busatta}, {Busso}, {Buzasi}, {Caballero}, {Cabral}, {Calderone}, {Cameron}, {Cameron}, {Campante}, {Canto Martins}, {Cara}, {Carone}, {Carrasco}, {Casagrande}, {Casewell}, {Cassisi}, {Castellani},
  {Castro}, {Catala}, {Catal{\'a}n Fern{\'a}ndez}, {Catelan}, {Cegla}, {Cerruti}, {Cessa}, {Chadid}, {Chaplin}, {Charpinet}, {Chiappini}, {Chiarucci}, {Chiavassa}, {Chinellato}, {Chirulli}, {Christensen-Dalsgaard}, {Church}, {Claret}, {Clarke}, {Claudi}, {Clermont}, {Coelho}, {Coelho}, {Cogato}, {Colom{\'e}}, {Condamin}, {Conseil}, {Corbard}, {Correia}, {Corsaro}, {Cosentino}, {Costes}, {Cottinelli}, {Covone}, {Creevey}, {Crida}, {Csizmadia}, {Cunha}, {Curry}, {da Costa}, {da Silva}, {Dalal}, {Damasso}, {Damiani}, {Damiani}, {Liduina das Chagas}, {Davies}, {Davies}, {Davies}, {Davison}, {de Almeida}, {de Angeli}, {Cabral de Barros}, {de Castro Le{\~a}o}, {Brito de Freitas}, {de Freitas}, {De Martino}, {Renan de Medeiros}, {de Paula}, {de Plaa}, {De Ridder}, {Deal}, {Decin}, {Deeg}, {Degl'Innocenti}, {Deheuvels}, {del Burgo}, {Del Sordo}, {Delgado-Mena}, {Demangeon}, {Denk}, {Derekas}, {Desidera}, {Dexet}, {Di Criscienzo}, {Di Giorgio}, {Di Mauro}, {Diaz Rial}, {D{\'\i}az-Garc{\'\i}a}, {Dima}, {Dinuzzi},
  {Dionatos}, {Distefano}, {do Nascimento}, {Domingo}, {D'Orazi}, {Dorn}, {Doyle}, {Duarte}, {Ducellier}, {Dumaye}, {Dumusque}, {Dupret}, {Eggenberger}, {Ehrenreich}, {Eigm{\"u}ller}, {Eising}, {Emilio}, {Eriksson}, {Ermocida}, {Isidoro Escate Giribaldi}, {Eschen}, {Estrela}, {Evans}, {Fabbian}, {Fabrizio}, {Faria}, {Farina}, {Farinato}, {Feliz}, {Feltzing}, {Fenouillet}, {Ferrari}, {Ferraz-Mello}, {Fialho}, {Fienga}, {Figueira}, {Fiori}, {Flaccomio}, {Focardi}, {Foley}, {Fontignie}, {Ford}, {Fornazier}, {Forveille}, {Fossati}, {de Marca Franca}, {da Silva}, {Frasca}, {Fridlund}, {Furlan}, {Gabler}, {Gaido}, {Gallagher}, {Galli}, {Garcia}, {Garc{\'\i}a Hern{\'a}ndez}, {Garcia Munoz}, {Garc{\'\i}a-V{\'a}zquez}, {Garrido Haba}, {Gaulme}, {Gauthier}, {Gehan}, {Gent}, {Georgieva}, {Ghigo}, {Giana}, {Gill}, {Girardi}, {Giuliatti Winter}, {Giusi}, {Gomes da Silva}, {G{\'o}mez Zazo}, {Gomez-Lopez}, {Isai Gonz{\'a}lez Hern{\'a}ndez}, {Gonzalez Murillo}, {Gorius}, {Gouel}, {Goulty}, {Granata}, {Grenfell},
  {Grie{\ss}bach}, {Grolleau}, {Grouffal}, {Grziwa}, {Guarcello}, {Gueguen}, {Guenther}, {Guilhem}, {Guillerot}, {Guiot}, {Guterman}, {Guti{\'e}rrez}, {Guti{\'e}rrez-Canales}, {Hagelberg}, {Haldemann}, {Hall}, {Handberg}, {Harrison}, {Harrison}, {Hasiba}, {Haswell}, {Hatalova}, {Hatzes}, {Haywood}, {H{\'e}brard}, {Heckes}, {Heiter}, {Hekker}, {Heller}, {Helling}, {Helminiak}, {Hemsley}, {Heng}, {Hermans}, {Hermes}, {Hidalgo Torres}, {Hinkel}, {Hobbs}, {Hodgkin}, {Hofmann}, {Hojjatpanah}, {Houdek}, {Huber}, {Huesler}, {Hui-Bon-Hoa}, {Huygen}, {Huynh}, {Iro}, {Irwin}, {Irwin}, {Izidoro}, {Jacquinod}, {Emborg Jannsen}, {Janson}, {Jeszenszky}, {Jiang}, {Jos{\'e} Jimenez Mancebo}, {Jofre}, {Johansen}, {Johnston}, {Jones}, {Kallinger}, {K{\'a}lm{\'a}n}, {Kanitz}, {Karjalainen}, {Karjalainen}, {Karoff}, {Kawaler}, {Kawata}, {Keereman}, {Keiderling}, {Kennedy}, {Kenworthy}, {Kerschbaum}, {Kidger}, {Kiefer}, {Kintziger}, {Kislyakova}, {Kiss}, {Klagyivik}, {Klahr}, {Klevas}, {Kochukhov}, {K{\"o}hler}, {Kolb}, {Koncz},
  {Korth}, {Kostogryz}, {Kov{\'a}cs}, {Kov{\'a}cs}, {Kozhura}, {Krivova}, {Ku{\v{c}}inskas}, {Kuhlemann}, {Kupka}, {Laauwen}, {Labiano}, {Lagarde}, {Laget}, {Laky}, {Lam}, {Lambrechts}, {Lammer}, {Lanza}, {Lanzafame}, {Lares Martiz}, {Laskar}, {Latter}, {Lavanant}, {Lawrenson}, {Lazzoni}, {Lebre}, {Lebreton}, {Lecavelier des Etangs}, {Leinhardt}, {Leleu}, {Lendl}, {Leto}, {Levillain}, {Libert}, {Lichtenberg}, {Ligi}, {Lignieres}, {Lillo-Box}, {Linsky}, {Scige Liu}, {Loidolt}, {Longval}, {Lopes}, {Lorenzani}, {Ludwig}, {Lund}, {Sloth Lundkvist}, {Luri}, {Maceroni}, {Madden}, {Madhusudhan}, {Maggio}, {Magliano}, {Magrin}, {Mahy}, {Maibaum}, {Malac-Allain}, {Malapert}, {Malavolta}, {Maldonado}, {Mamonova}, {Manchon}, {Mann}, {Mantovan}, {Marafatto}, {Marconi}, {Mardling}, {Marigo}, {Marinoni}, {Marques}, {Marques}, {Marrese}, {Marshall}, {Mart{\'\i}nez Perales}, {Mary}, {Marzari}, {Masana}, {Mascher}, {Mathis}, {Mathur}, {Mattiuci Figueiredo}, {Maxted}, {Mazeh}, {Mazevet}, {Mazzei}, {McCormac}, {McMillan},
  {Menou}, {Merle}, {Meru}, {Mesa}, {Messina}, {M{\'e}sz{\'a}ros}, {Meunier}, {Meunier}, {Micela}, {Michaelis}, {Michel}, {Michielsen}, {Michtchenko}, {Miglio}, {Miguel}, {Milligan}, {Mirouh}, {Mitchel}, {Moedas}, {Molendini}, {Moln{\'a}r}, {Mombarg}, {Montalban}, {Montalto}, {Monteiro}, {Morales}, {Morales-Calderon}, {Morbidelli}, {Mordasini}, {Moreau}, {Morel}, {Morello}, {Morin}, {Mortier}, {Mosser}, {Mourard}, {Mousis}, {Moutou}, {Mowlavi}, {Moya}, {Muehlmann}, {Muirhead}, {Munari}, {Musella}, {Mustill}, {Nardetto}, {Nardiello}, {Narita}, {Nascimbeni}, {Nash}, {Neiner}, {Nelson}, {Nettelmann}, {Nicolini}, {Nielsen}, {Niemi}, {Noack}, {Noels-Grotsch}, {Noll}, {Norazman}, {Norton}, {Nsamba}, {Ofir}, {Ogilvie}, {Olander}, {Olivetto}, {Olofsson}, {Ong}, {Ortolani}, {Oshagh}, {Ottacher}, {Ottensamer}, {Ouazzani}, {Paardekooper}, {Pace}, {Pajas}, {Palacios}, {Palandri}, {Palle}, {Paproth}, {Parro}, {Parviainen}, {Granado}, {Passegger}, {Pastor-Morales}, {P{\"a}tzold}, {Gade Pedersen}, {Pena Hidalgo}, {Pepe},
  {Pereira}, {Persson}, {Pertenais}, {Peter}, {Petit}, {Petit}, {Pezzuto}, {Pichierri}, {Pietrinferni}, {Pinheiro}, {Pinsonneault}, {Plachy}, {Plasson}, {Plez}, {Poppenhaeger}, {Poretti}, {Portaluri}, {Portell}, {Frederico Porto de Mello}, {Poyatos}, {Pozuelos}, {Prada Moroni}, {Pricopi}, {Prisinzano}, {Quade}, {Quirrenbach160}, {Rabanal Reina6}, {Rabello Soares}, {Raimondo}, {Rainer}, {Ram{\'o}n Rod{\'o}n}, {Ram{\'o}n-Ballesta}, {Ramos Zapata}, {R{\"a}tz}, {Rauterberg}, {Redman}, {Redmer}, {Reese}, {Regibo}, {Reiners}, {Reinhold}, {Renie}, {Ribas}, {Ribeiro}, {Pereira Ricciardi}, {Rice}, {Richard}, {Riello}, {Rieutord}, {Ripepi}, {Rixon}, {Rockstein}, {Rodr{\'\i}guez}, {Rodr{\'\i}guez D{\'\i}az}, {Rodriguez Garcia}, {Rodriguez-Gomez}, {Roehlly}, {Roig}, {Rojas-Ayala}, {Rolf}, {Lysgaard R{\o}rsted}, {Rosado}, {Rosotti}, {Roth}, {Roth}, {Rousseau}, {Roxburgh}, {Roy}, {Royer}, {Ruane}, {Rufini Mastropasqua}, {Ruiz de Galarreta}, {Russi}, {Saar}, {Saillenfest}, {Salaris}, {Salmon}, {Saltas}, {Samadi}, {Samadi},
  {Samra}, {Sanches da Silva}, {Andr{\'e}s S{\'a}nchez Carrasco}, {Santerne}, {Santoli}, {Santos}, {Sanz Mesa}, {Sarro}, {Scandariato}, {Sch{\"a}fer}, {Schlafly}, {Schmider}, {Schneider}, {Schou}, {Schunker}, {J{\"o}rg Schwarzkopf}, {Serenelli}, {Seynaeve}, {Shan}, {Shapiro}, {Shipman}, {Sicilia}, {Sierra Sanmartin}, {Sigot}, {Silliman}, {Silvotti}, {Simon}, {Simoyama Napoli}, {Skarka}, {Smalley}, {Smiljanic}, {Smit}, {Smith}, {Smith}, {Snellen}, {S{\'o}dor}, {Sohl}, {Solanki}, {Sortino}, {Sousa}, {Southworth}, {Souto}, {Sozzetti}, {Stamatellos}, {Stassun}, {Steller}, {Stello}, {Stelzer}, {Stiebeler}, {Stokholm}, {Storelvmo}, {Strassmeier}, {Str{\o}m}, {Strugarek}, {Sulis}, {{\v{S}}vanda}, {Szabados}, {Szab{\'o}}, {Szab{\'o}}, {Szuszkiewicz}, {Talens}, {Teti}, {Theisen}, {Th{\'e}venin}, {Thoul}, {Tiphene}, {Titz-Weider}, {Tkachenko}, {Tomecki}, {Tonfat}, {Tosi}, {Trampedach}, {Traven}, {Triaud}, {Tr{\o}nnes}, {Tsantaki}, {Tschentscher}, {Turin}, {Tvaruzka}, {Ulmer}, {Ulmer-Moll}, {Ulusoy}, {Umbriaco},
  {Valencia}, {Valentini}, {Valio}, {Valverde Guijarro}, {Van Eylen}, {Van Grootel}, {van Kempen}, {Van Reeth}, {Van Zelst}, {Vandenbussche}, {Vasiliou}, {Vasilyev}, {Vaz de Mascarenhas}, {Vazan}, {Vela Nunez}, {Nunes Velloso}, {Ventura}, {Ventura}, {Venturini}, {Trallero}, {Veras}, {Verdugo}, {Verma}, {Vibert}, {Vicanek Martinez}, {Vida}, {Vigan}, {Villacorta}, {Villaver}, {Villaverde Aparicio}, {Viotto}, {Vorobyov}, {Vorontsov}, {Wagner}, {Walloschek}, {Walton}, {Walton}, {Wang}, {Waters}, {Watson}, {Wedemeyer}, {Weeks}, {Weingril}, {Weiss}, {Wendler}, {West}, {Westerdorff}, {Westphal}, {Wheatley}, {White}, {Whittaker}, {Wickhusen}, {Wilson}, {Windsor}, {Winter}, {Lykke Winther}, {Winton}, {Witteck}, {Witzke}, {Woitke}, {Wolter}, {Wuchterl}, {Wyatt}, {Yang}, {Yu}, {Zanmar Sanchez}, {Rosa Zapatero Osorio}, {Zechmeister}, {Zhou}, {Ziemke}, \& {Zwintz}}]{plato_ref}
{Rauer}, H., {Aerts}, C., {Cabrera}, J., {et~al.} 2024, arXiv e-prints, arXiv:2406.05447

\bibitem[{{Rogers} {et~al.}(2023){Rogers}, {Schlichting}, \& {Owen}}]{Rogers23}
{Rogers}, J.~G., {Schlichting}, H.~E., \& {Owen}, J.~E. 2023, \apjl, 947, L19

\bibitem[{{Sainsbury-Martinez} {et~al.}(2019){Sainsbury-Martinez}, {Wang}, {Fromang}, {Tremblin}, {Dubos}, {Meurdesoif}, {Spiga}, {Leconte}, {Baraffe}, {Chabrier}, {Mayne}, {Drummond}, \& {Debras}}]{SMartinez19}
{Sainsbury-Martinez}, F., {Wang}, P., {Fromang}, S., {et~al.} 2019, \aap, 632, A114

\bibitem[{{Sarkis} {et~al.}(2021){Sarkis}, {Mordasini}, {Henning}, {Marleau}, \& {Molli{\`e}re}}]{Sarkis21}
{Sarkis}, P., {Mordasini}, C., {Henning}, T., {Marleau}, G.~D., \& {Molli{\`e}re}, P. 2021, \aap, 645, A79

\bibitem[{{Saumon} {et~al.}(1995){Saumon}, {Chabrier}, \& {van Horn}}]{Saumon95}
{Saumon}, D., {Chabrier}, G., \& {van Horn}, H.~M. 1995, \apjs, 99, 713

\bibitem[{{Schneider} {et~al.}(2022){Schneider}, {Carone}, {Decin}, {J{\o}rgensen}, \& {Helling}}]{Schneider22}
{Schneider}, A.~D., {Carone}, L., {Decin}, L., {J{\o}rgensen}, U.~G., \& {Helling}, C. 2022, \aap, 666, L11

\bibitem[{{Seiff} {et~al.}(1998){Seiff}, {Kirk}, {Knight}, {Young}, {Mihalov}, {Young}, {Milos}, {Schubert}, {Blanchard}, \& {Atkinson}}]{Seiff98}
{Seiff}, A., {Kirk}, D.~B., {Knight}, T. C.~D., {et~al.} 1998, \jgr, 103, 22857

\bibitem[{{Showman} \& {Guillot}(2002)}]{Showman02}
{Showman}, A.~P. \& {Guillot}, T. 2002, \aap, 385, 166

\bibitem[{{Showman} \& {Polvani}(2011)}]{Showman11}
{Showman}, A.~P. \& {Polvani}, L.~M. 2011, \apj, 738, 71

\bibitem[{{Sing} {et~al.}(2024){Sing}, {Rustamkulov}, {Thorngren}, {Barstow}, {Tremblin}, {Alves de Oliveira}, {Beck}, {Birkmann}, {Challener}, {Crouzet}, {Espinoza}, {Ferruit}, {Giardino}, {Gressier}, {Lee}, {Lewis}, {Maiolino}, {Manjavacas}, {Rauscher}, {Sirianni}, \& {Valenti}}]{Sing24}
{Sing}, D.~K., {Rustamkulov}, Z., {Thorngren}, D.~P., {et~al.} 2024, arXiv e-prints, arXiv:2405.11027

\bibitem[{{Spiegel} \& {Burrows}(2012)}]{Spiegel12}
{Spiegel}, D.~S. \& {Burrows}, A. 2012, \apj, 745, 174

\bibitem[{{Stevenson} {et~al.}(2016){Stevenson}, {Bean}, {Seifahrt}, {Gilbert}, {Line}, {D{\'e}sert}, \& {Fortney}}]{Stevenson16}
{Stevenson}, K.~B., {Bean}, J.~L., {Seifahrt}, A., {et~al.} 2016, \apj, 817, 141

\bibitem[{{Sur} {et~al.}(2024){Sur}, {Su}, {Tejada Arevalo}, {Chen}, \& {Burrows}}]{apple_model}
{Sur}, A., {Su}, Y., {Tejada Arevalo}, R., {Chen}, Y.-X., \& {Burrows}, A. 2024, arXiv e-prints, arXiv:2404.14483

\bibitem[{{Teske} {et~al.}(2019){Teske}, {Thorngren}, {Fortney}, {Hinkel}, \& {Brewer}}]{Teske19}
{Teske}, J.~K., {Thorngren}, D., {Fortney}, J.~J., {Hinkel}, N., \& {Brewer}, J.~M. 2019, \aj, 158, 239

\bibitem[{{Thorngren} \& {Fortney}(2018)}]{Thorngren18}
{Thorngren}, D.~P. \& {Fortney}, J.~J. 2018, \aj, 155, 214

\bibitem[{{Thorngren} {et~al.}(2016){Thorngren}, {Fortney}, {Murray-Clay}, \& {Lopez}}]{Thorngren16}
{Thorngren}, D.~P., {Fortney}, J.~J., {Murray-Clay}, R.~A., \& {Lopez}, E.~D. 2016, \apj, 831, 64

\bibitem[{{Tinetti} {et~al.}(2018){Tinetti}, {Drossart}, {Eccleston}, {Hartogh}, {Heske}, {Leconte}, {Micela}, {Ollivier}, {Pilbratt}, {Puig}, {Turrini}, {Vandenbussche}, {Wolkenberg}, {Beaulieu}, {Buchave}, {Ferus}, {Griffin}, {Guedel}, {Justtanont}, {Lagage}, {Machado}, {Malaguti}, {Min}, {N{\o}rgaard-Nielsen}, {Rataj}, {Ray}, {Ribas}, {Swain}, {Szabo}, {Werner}, {Barstow}, {Burleigh}, {Cho}, {Coud{\'e} du Foresto}, {Coustenis}, {Decin}, {Encrenaz}, {Galand}, {Gillon}, {Helled}, {Morales}, {Garc{\'\i}a Mu{\~n}oz}, {Moneti}, {Pagano}, {Pascale}, {Piccioni}, {Pinfield}, {Sarkar}, {Selsis}, {Tennyson}, {Triaud}, {Venot}, {Waldmann}, {Waltham}, {Wright}, {Amiaux}, {Augu{\`e}res}, {Berth{\'e}}, {Bezawada}, {Bishop}, {Bowles}, {Coffey}, {Colom{\'e}}, {Crook}, {Crouzet}, {Da Peppo}, {Sanz}, {Focardi}, {Frericks}, {Hunt}, {Kohley}, {Middleton}, {Morgante}, {Ottensamer}, {Pace}, {Pearson}, {Stamper}, {Symonds}, {Rengel}, {Renotte}, {Ade}, {Affer}, {Alard}, {Allard}, {Altieri}, {Andr{\'e}}, {Arena}, {Argyriou},
  {Aylward}, {Baccani}, {Bakos}, {Banaszkiewicz}, {Barlow}, {Batista}, {Bellucci}, {Benatti}, {Bernardi}, {B{\'e}zard}, {Blecka}, {Bolmont}, {Bonfond}, {Bonito}, {Bonomo}, {Brucato}, {Brun}, {Bryson}, {Bujwan}, {Casewell}, {Charnay}, {Pestellini}, {Chen}, {Ciaravella}, {Claudi}, {Cl{\'e}dassou}, {Damasso}, {Damiano}, {Danielski}, {Deroo}, {Di Giorgio}, {Dominik}, {Doublier}, {Doyle}, {Doyon}, {Drummond}, {Duong}, {Eales}, {Edwards}, {Farina}, {Flaccomio}, {Fletcher}, {Forget}, {Fossey}, {Fr{\"a}nz}, {Fujii}, {Garc{\'\i}a-Piquer}, {Gear}, {Geoffray}, {G{\'e}rard}, {Gesa}, {Gomez}, {Graczyk}, {Griffith}, {Grodent}, {Guarcello}, {Gustin}, {Hamano}, {Hargrave}, {Hello}, {Heng}, {Herrero}, {Hornstrup}, {Hubert}, {Ida}, {Ikoma}, {Iro}, {Irwin}, {Jarchow}, {Jaubert}, {Jones}, {Julien}, {Kameda}, {Kerschbaum}, {Kervella}, {Koskinen}, {Krijger}, {Krupp}, {Lafarga}, {Landini}, {Lellouch}, {Leto}, {Luntzer}, {Rank-L{\"u}ftinger}, {Maggio}, {Maldonado}, {Maillard}, {Mall}, {Marquette}, {Mathis}, {Maxted}, {Matsuo},
  {Medvedev}, {Miguel}, {Minier}, {Morello}, {Mura}, {Narita}, {Nascimbeni}, {Nguyen Tong}, {Noce}, {Oliva}, {Palle}, {Palmer}, {Pancrazzi}, {Papageorgiou}, {Parmentier}, {Perger}, {Petralia}, {Pezzuto}, {Pierrehumbert}, {Pillitteri}, {Piotto}, {Pisano}, {Prisinzano}, {Radioti}, {R{\'e}ess}, {Rezac}, {Rocchetto}, {Rosich}, {Sanna}, {Santerne}, {Savini}, {Scandariato}, {Sicardy}, {Sierra}, {Sindoni}, {Skup}, {Snellen}, {Sobiecki}, {Soret}, {Sozzetti}, {Stiepen}, {Strugarek}, {Taylor}, {Taylor}, {Terenzi}, {Tessenyi}, {Tsiaras}, {Tucker}, {Valencia}, {Vasisht}, {Vazan}, {Vilardell}, {Vinatier}, {Viti}, {Waters}, {Wawer}, {Wawrzaszek}, {Whitworth}, {Yung}, {Yurchenko}, {Zapatero Osorio}, {Zellem}, {Zingales}, \& {Zwart}}]{ariel_ref}
{Tinetti}, G., {Drossart}, P., {Eccleston}, P., {et~al.} 2018, Experimental Astronomy, 46, 135

\bibitem[{{Tremblin} {et~al.}(2017){Tremblin}, {Chabrier}, {Mayne}, {Amundsen}, {Baraffe}, {Debras}, {Drummond}, {Manners}, \& {Fromang}}]{Tremblin17}
{Tremblin}, P., {Chabrier}, G., {Mayne}, N.~J., {et~al.} 2017, \apj, 841, 30

\bibitem[{{Tsai} {et~al.}(2021){Tsai}, {Malik}, {Kitzmann}, {Lyons}, {Fateev}, {Lee}, \& {Heng}}]{vulcan_21}
{Tsai}, S.-M., {Malik}, M., {Kitzmann}, D., {et~al.} 2021, \apj, 923, 264

\bibitem[{{Vazan} {et~al.}(2018){Vazan}, {Helled}, \& {Guillot}}]{Vazan18}
{Vazan}, A., {Helled}, R., \& {Guillot}, T. 2018, \aap, 610, L14

\bibitem[{{Vazan} {et~al.}(2015){Vazan}, {Helled}, {Kovetz}, \& {Podolak}}]{Vazan15}
{Vazan}, A., {Helled}, R., {Kovetz}, A., \& {Podolak}, M. 2015, \apj, 803, 32

\bibitem[{{Vazan} {et~al.}(2022){Vazan}, {Sari}, \& {Kessel}}]{Vazan22}
{Vazan}, A., {Sari}, R., \& {Kessel}, R. 2022, \apj, 926, 150

\bibitem[{{Virtanen} {et~al.}(2020){Virtanen}, {Gommers}, {Oliphant}, {Haberland}, {Reddy}, {Cournapeau}, {Burovski}, {Peterson}, {Weckesser}, {Bright}, {van der Walt}, {Brett}, {Wilson}, {Millman}, {Mayorov}, {Nelson}, {Jones}, {Kern}, {Larson}, {Carey}, {Polat}, {Feng}, {Moore}, {VanderPlas}, {Laxalde}, {Perktold}, {Cimrman}, {Henriksen}, {Quintero}, {Harris}, {Archibald}, {Ribeiro}, {Pedregosa}, {van Mulbregt}, \& {SciPy 1. 0 Contributors}}]{Virtanen20}
{Virtanen}, P., {Gommers}, R., {Oliphant}, T.~E., {et~al.} 2020, Nature Methods, 17, 261

\bibitem[{{Visscher} \& {Moses}(2011)}]{Visscher11}
{Visscher}, C. \& {Moses}, J.~I. 2011, \apj, 738, 72

\bibitem[{{Wagner} \& {Pru{\ss}}(2002)}]{Wagner02}
{Wagner}, W. \& {Pru{\ss}}, A. 2002, Journal of Physical and Chemical Reference Data, 31, 387

\bibitem[{{Wahl} {et~al.}(2017){Wahl}, {Hubbard}, {Militzer}, {Guillot}, {Miguel}, {Movshovitz}, {Kaspi}, {Helled}, {Reese}, {Galanti}, {Levin}, {Connerney}, \& {Bolton}}]{Wahl17}
{Wahl}, S.~M., {Hubbard}, W.~B., {Militzer}, B., {et~al.} 2017, \grl, 44, 4649

\bibitem[{{Wakeford} {et~al.}(2017){Wakeford}, {Sing}, {Kataria}, {Deming}, {Nikolov}, {Lopez}, {Tremblin}, {Amundsen}, {Lewis}, {Mandell}, {Fortney}, {Knutson}, {Benneke}, \& {Evans}}]{Wakeford17}
{Wakeford}, H.~R., {Sing}, D.~K., {Kataria}, T., {et~al.} 2017, Science, 356, 628

\bibitem[{{Welbanks} {et~al.}(2024){Welbanks}, {Bell}, {Beatty}, {Line}, {Ohno}, {Fortney}, {Schlawin}, {Greene}, {Rauscher}, {McGill}, {Murphy}, {Parmentier}, {Tang}, {Edelman}, {Mukherjee}, {Wiser}, {Lagage}, {Dyrek}, \& {Arnold}}]{Welbanks24}
{Welbanks}, L., {Bell}, T.~J., {Beatty}, T.~G., {et~al.} 2024, arXiv e-prints, arXiv:2405.11018

\bibitem[{{Zahnle} \& {Marley}(2014)}]{Zahnle14}
{Zahnle}, K.~J. \& {Marley}, M.~S. 2014, \apj, 797, 41

\bibitem[{{Zeng} {et~al.}(2019){Zeng}, {Jacobsen}, {Sasselov}, {Petaev}, {Vanderburg}, {Lopez-Morales}, {Perez-Mercader}, {Mattsson}, {Li}, {Heising}, {Bonomo}, {Damasso}, {Berger}, {Cao}, {Levi}, \& {Wordsworth}}]{Zeng19}
{Zeng}, L., {Jacobsen}, S.~B., {Sasselov}, D.~D., {et~al.} 2019, Proceedings of the National Academy of Science, 116, 9723

\bibitem[{{Zhang} {et~al.}(2023){Zhang}, {Li}, {Ge}, \& {Le}}]{Zhang23}
{Zhang}, X., {Li}, C., {Ge}, H., \& {Le}, T. 2023, \apj, 957, 22

\end{thebibliography}

\end{document}